\documentclass[aps,preprint,floatfix,nofootinbib,showpacs]{revtex4-1}

\usepackage{graphicx,color}
\usepackage{epsfig}

\newcommand{\lsim}{\raisebox{-0.13cm}{~\shortstack{$<$ \\[-0.07cm] $\sim$}}~}
\newcommand{\gsim}{\raisebox{-0.13cm}{~\shortstack{$>$ \\[-0.07cm] $\sim$}}~}

\begin{document}

{\small
\begin{flushright}
CNU-HEP-14-02
\end{flushright} }

\title{Confronting Higgcision with Electric Dipole Moments}

\renewcommand{\thefootnote}{\arabic{footnote}}

\author{
Kingman Cheung$^{1,2}$, Jae Sik Lee$^3$, 
Eibun Senaha$^4$, and Po-Yan Tseng$^1$}

\affiliation{ 
$^1$ Department of Physics, National Tsing Hua University,
Hsinchu 300, Taiwan \\
$^2$ Division of Quantum Phases and Devices, School of Physics, 
Konkuk University, Seoul 143-701, Republic of Korea \\
$^3$ Department of Physics, Chonnam National University, \\
300 Yongbong-dong, Buk-gu, Gwangju, 500-757, Republic of Korea \\
$^4$
Department of Physics, Nagoya University, Nagoya 464-8602, Japan
}
\date{March 19, 2014}

\begin{abstract}
Current data on the signal strengths and angular spectrum of the 125.5
GeV Higgs boson still allow a CP-mixed state, namely, the pseudoscalar
coupling to the top quark can be as sizable as the scalar coupling:
$C_u^S \approx C_u^P =1/2$.  CP violation can then arise and manifest
in sizable electric dipole moments (EDMs).  In the framework of
two-Higgs-doublet models, we not only update the Higgs precision 
(Higgcision) study on the couplings with the most updated Higgs signal
strength data, but also compute all the Higgs-mediated contributions
from the 125.5 GeV Higgs boson to the EDMs, and confront the allowed
parameter space against the existing constraints from the EDM
measurements of Thallium, neutron, Mercury, and Thorium monoxide.  We
found that the combined EDM constraints restrict the pseudoscalar
coupling to be less than about $10^{-2}$,
unless there are contributions
from other Higgs bosons, supersymmetric particles, or other exotic
particles that delicately cancel the current Higgs-mediated contributions.
\end{abstract}

\maketitle

\section{Introduction}
Since the observation of a new boson 
at a mass around $125.5$ GeV at the Large Hadron Collider (LHC) 
\cite{atlas,cms}, the most urgent mission is to investigate
the properties of this new boson.
There have been a large number of studies or fits of the Higgs boson
couplings to the standard model (SM) particles in
more or less model-independent frameworks~\cite{r1,
r2,r3,r4,r5,r6,r7,r8,r9,r10,r11,r12,r13,r14,r15,r16,r17,r18,Cheung:2013kla,
r19,r20,r21,r22,anom1,anom2,anom3,anom4,anom5,anom6}, 
in the two-Higgs doublet model (2HDM) frameworks~\cite{2hdm0,2hdm1,2hdm2,
2hdm3,2hdm4,2hdm5,2hdm6,2hdm7,2hdm8,2hdm9,2hdm10,2hdm11,
2hdm12,2hdm13,2hdm14,2hdm15,2hdm16,2hdm17,Cheung:2013rva}, and in the
supersymmetric frameworks~\cite{susy1,susy2,susy3,susy4,susy5}.
Based on a study using a generic framework for Higgs couplings to
the relevant SM particles, three of us has 
reported~\cite{Cheung:2013kla} 
that the SM Higgs boson~\cite{higgs} provides the best fit to all the
most updated Higgs data from
ATLAS~\cite{atlas_h_aa_2013,atlas_com_2013,atlas_bb_2013,atlas_h_tau_2013},
CMS~\cite{cms_aa_2013,cms_zz_2013,cms_ww_2013,cms_tau_2013,cms_tau_2013_update,
cms_bb_vh,cms_bb_tth},
and Tevatron~\cite{tev,tev_h_bb}. 
In particular, the relative coupling to the gauge bosons is restricted
to be close to the SM values with about a 15\% uncertainty while
the Yukawa couplings are only loosely constrained.
Furthermore, the hypothesis of a pure CP-odd state for the new boson 
has been mostly ruled out by angular 
measurements~\cite{Chatrchyan:2012jja,Aad:2013xqa}.
Nevertheless, there is still a large room
for the possibility of a CP-mixed state~\cite{r16-1,Cheung:2013kla}.

If the Higgs boson is a CP-mixed state, it can 
simultaneously couple to the scalar and pseudoscalar fermion bilinears
as follows:
\begin{equation}
{\cal L}_{H\bar{f}f}\ =\ - \,g_f\,
H\, \bar{f}\,\Big( g^S_{H\bar{f}f}\, +\,
ig^P_{H\bar{f}f}\gamma_5 \Big)\, f\ ,
\end{equation}
where $g_f=g m_f/2 M_W=m_f/v$ with $f=u,d,l$ 
denoting the up- and down-type quarks
and charged leptons collectively. 
We will show that non-zero values of the products proportional to 
$g^S_{H\bar{f}f}\times g^P_{H\bar{f}^{(\prime)}f^{(\prime)}}$ 
and $g^P_{H\bar{f}f}\times g_{HVV}$ signal CP violation as manifested
in nonzero values for electric dipole moments (EDMs)
\footnote{Here,
$g_{HVV}$ denotes a generic Higgs coupling to the massive vector bosons
in the interaction ${\cal L}_{HVV}  =  g\,M_W \, g_{HVV}\,\left(
 W^+_\mu W^{- \mu} +  \frac{1}{2\cos^2\theta_W}\,Z_\mu Z^\mu\right) \, H$.}.
The non-observation  of  the 
Thallium ($^{205}{\rm Tl}$)~\cite{Regan:2002ta},    
neutron~($n$)~\cite{Baker:2006ts}, 
Mercury ($^{199}{\rm Hg}$)~\cite{Griffith:2009zz}, 
and thorium monoxide (ThO)~\cite{Baron:2013eja} EDMs
provide remarkably  tight bounds on  CP violation. 
The EDM constraints in light of the recent Higgs data were studied in
Refs.~\cite{Shu:2013uua,Brod:2013cka}.
Strictly speaking, only the Higgs couplings to the third-generation
fermions such as the top and bottom quarks and tau leptons are relevant to
the current Higgs data.  On the other hand, the EDM
experiments mainly involve the first-generation fermions.
Therefore, it is impossible to relate the 
Higgs precision (Higgcision)
constraints to EDMs in a completely model-independent fashion without 
specifying the relations among the generations, 
except for the Weinberg operator.
In most of the models studied in literature, 
however, the Higgs couplings to the third-generation
fermions are related to those of the first-generation in a model-dependent way.
In this work, to be specific, we study the contributions
of the observed 125.5 GeV ``Higgs'' boson ($H$) to EDMs in the 
framework of 2HDMs.

The paper is organized as follows. In Sec.~II we briefly describe
ingredients in the framework of 2HDMs we are working with and present 
the 2HDM Higgcision fit to the most updated Higgs data.
For notation and more details of the 2HDMs we refer to 
Ref.~\cite{Cheung:2013rva}.
Section~III is devoted to the synopsis of EDMs.
In Sec.~IV we present our numerical results, and 
summarize our findings and draw conclusions in Sec.~V.

\section{Two Higgs Doublet Models}
In Ref.~\cite{Cheung:2013rva}, neglecting the charged Higgs contribution to
the loop-induced Higgs couplings to two photons,
it was shown that the Higgcision studies in 2HDM framework 
can be performed with a minimum of three parameters, given by
\begin{equation}
C_u^S\equiv g^S_{H\bar{t}t}\,; \ \ \
C_u^P\equiv g^P_{H\bar{t}t}\,; \ \ \
C_v\equiv g_{HVV}\,,
\end{equation}
where $H=h_i$ denotes the candidate of the 125.5 GeV Higgs among the 
three neutral Higgs bosons $h_{1,2,3}$ in 2HDMs without further 
specifying which one the observed one is.
The mixing between the mass eigenstates $h_{1,2,3}$ and 
the electroweak eigenstates $\phi_1,\phi_2,a$ 
is described  by an orthogonal matrix $O$ as in
\begin{eqnarray}
(\phi_1,\phi_2,a)^T_\alpha = O_{\alpha j} (h_1,h_2,h_3)^T_j\,.
\end{eqnarray}

\begin{table}[hb!]
\caption{\label{tab:cdcl}
{\it  The couplings $C_{d,l}^{S,P}\equiv g^{S,P}_{H\bar{d}d,H\bar{l}l}$ as
functions of $C_u^{S,P}$ and $\tan\beta$ in the four types of 2HDMs,
see Ref.~\cite{Cheung:2013rva} for details of conventions in 2HDMs.
}}
\begin{center}
\begin{tabular}{||l||c|c||c|c||}
\hline
2HDM I & $C_d^S = C_u^S$ & $C_l^S = C_u^S$ &
$C_d^P=-C_u^P$ & $C_l^P=-C_u^P$ \\
\hline
2HDM II & $C_d^S = \frac{O_{\phi_1 i}}{c_\beta}$ &
$C_l^S = \frac{O_{\phi_1 i}}{c_\beta}$ &
$C_d^P=t_\beta^2C_u^P$ & $C_l^P=t_\beta^2C_u^P$ \\
\hline
2HDM III & $C_d^S = C_u^S$ &
$C_l^S = \frac{O_{\phi_1 i}}{c_\beta}$ &
$C_d^P=-C_u^P$ & $C_l^P=t_\beta^2C_u^P$ \\
\hline
2HDM IV &
$C_d^S = \frac{O_{\phi_1 i}}{c_\beta}$ &
$C_l^S = C_u^S$ &
$C_d^P=t_\beta^2C_u^P$ & $C_l^P=-C_u^P$ \\
\hline
\end{tabular}
\end{center}
\end{table}

Once the three parameters $C_u^S$, $C_u^P$, and $C_v$ are given, 
the $H$ couplings to the SM
fermions are completely determined as shown in  Table \ref{tab:cdcl}
\footnote{One may use $\tan\beta$ as an input parameter instead of $C_v$. 
Then, the coupling $C_v$
is given by
$C_v=c_\beta O_{\phi_1 i} + s_\beta O_{\phi_2 i}$.}.
Note the relations
\begin{equation}
O_{\phi_1 i} = \pm \left[1-(O_{\phi_2 i})^2-(O_{a i})^2\right]^{1/2}\,, \ \ \
O_{\phi_2 i}=s_\beta\,C_u^S\,, \ \ \ 
O_{ai}=-t_\beta\,C_u^P 
\end{equation}
with
\begin{equation}
\label{eq:sbsq}
s_\beta^2
=\frac{(1-C_v^2)}{(1-C_v^2)+(C_u^S-C_v)^2+(C_u^P)^2}\,.
\end{equation}
We are using the abbreviations:
$s_\beta\equiv\sin\beta$, $c_\beta\equiv\cos\beta$,
$t_\beta=\tan\beta$, etc, and the convention of $C_v>0$.

\begin{figure}[th]
\hspace{ 0.0cm}
\vspace{-0.5cm}
\includegraphics[angle=-90,width=7cm]{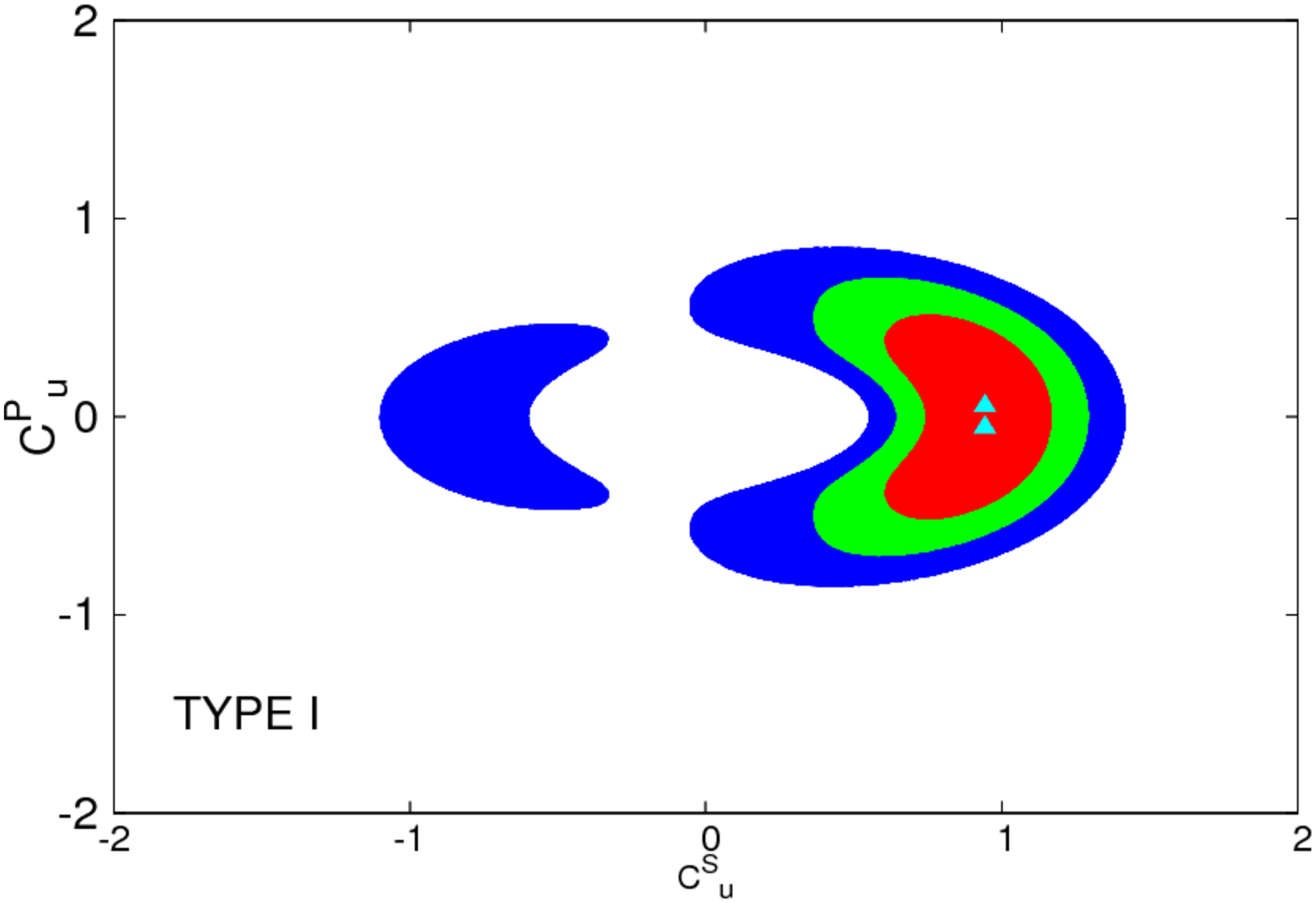}
\includegraphics[angle=-90,width=7cm]{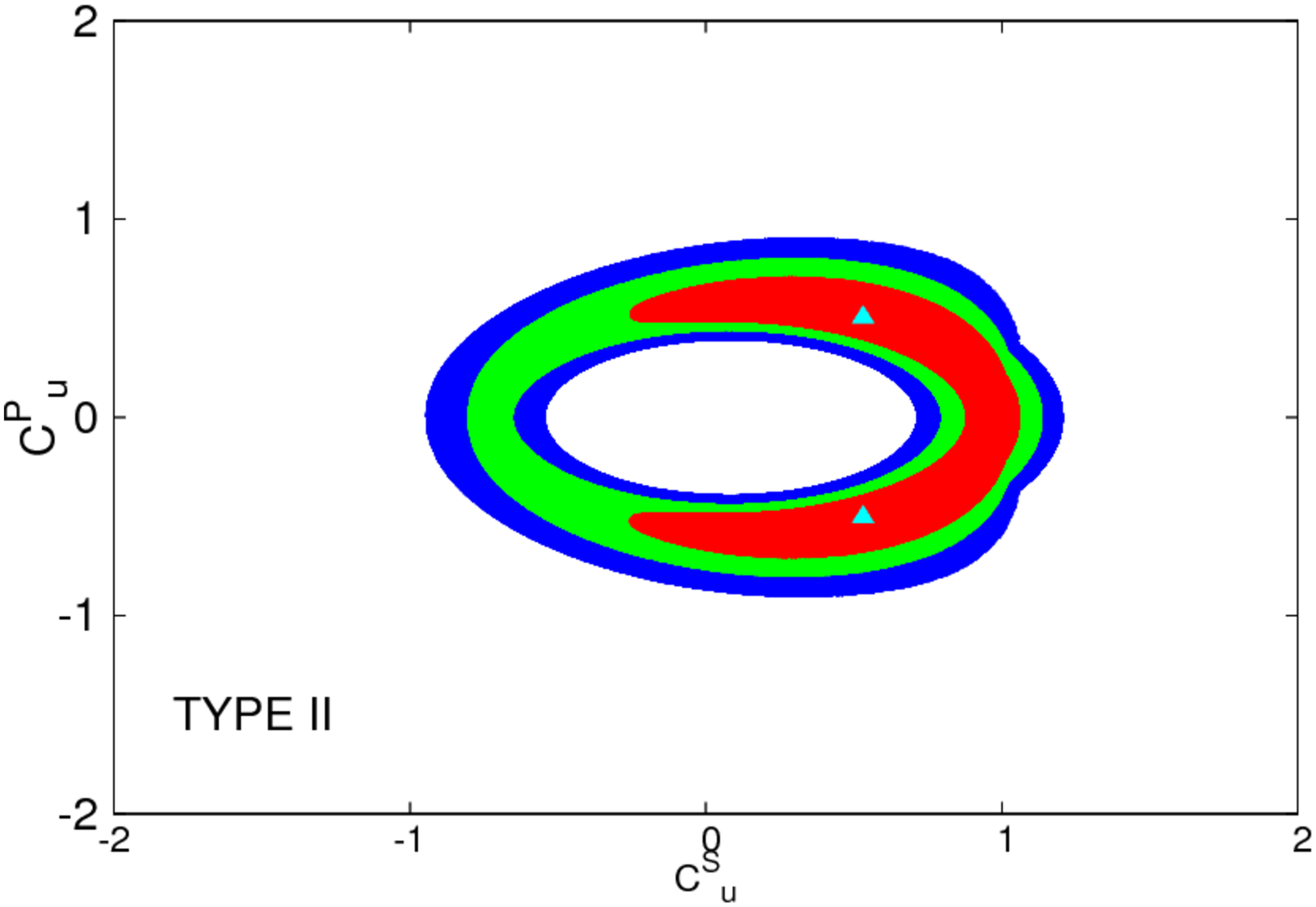}
\\[10mm]
\includegraphics[angle=-90,width=7cm]{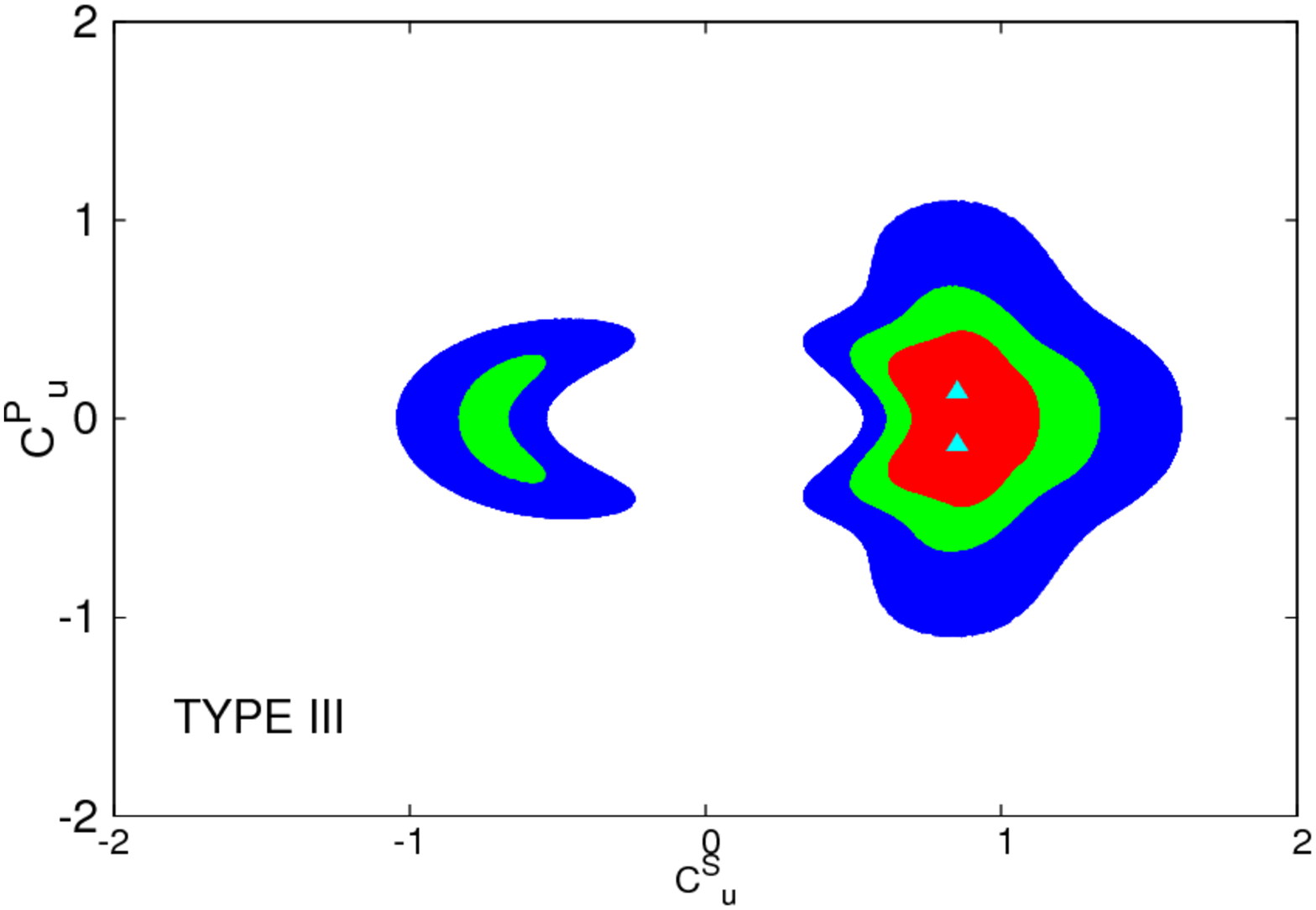}
\includegraphics[angle=-90,width=7cm]{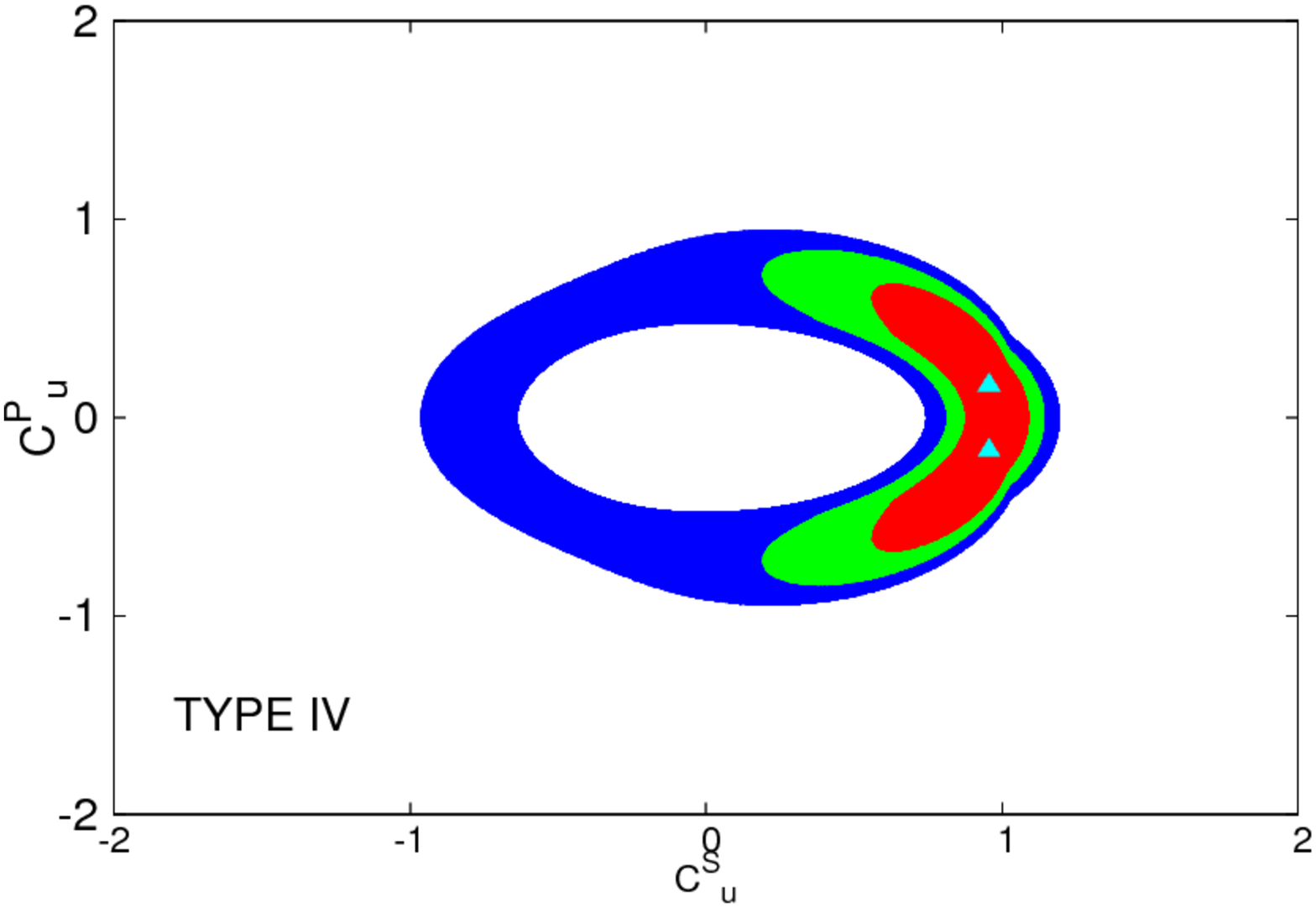}
\vspace{0.5cm}
\caption{\it The confidence-level regions of the fit
to the most updated Higgs data 
by varying $C_u^S$, $C_u^P$, and
$C_v$ in the plane of $C_u^S$ vs $C_u^P$ for Type I -- IV.
The contour regions shown are for
$\Delta \chi^2 \le 2.3$ (red), $5.99$ (green), and $11.83$ (blue)
above the minimum, which
correspond to confidence levels of
$68.3\%$, $95\%$, and $99.7\%$, respectively.
The best-fit points are denoted by the triangles.
}
\label{fig:hfit}
\end{figure}

In Fig.~\ref{fig:hfit}, we show the confidence-level (CL) regions of the fit
to the most updated Higgs data 
by varying $C_u^S$, $C_u^P$, and
$C_v$ in the plane of $C_u^S$ vs $C_u^P$ for Type I -- IV of the 2HDMs. 
Comparing to Fig.~11 in Ref.~\cite{Cheung:2013rva} for the {\bf CPV3} fit,
the CL regions are mildly reduced, 
preferring positive $C_u^S$ values slightly more than the negative ones, 
after the  inclusion of the most recent results from 
$H\to b\bar{b}$~\cite{atlas_bb_2013,cms_bb_vh,cms_bb_tth} 
and $\tau^+\tau^-$~\cite{atlas_h_tau_2013,cms_tau_2013_update}.
Meanwhile, we note that the maximal CP violation with
$C_u^S\sim |C_u^P|$ is still possible.
%

\section{Synopsis of EDMs}
Here we closely follow the methods used in 
Refs.~\cite{Ellis:2008zy,Ellis:2010xm,Ellis:2011hp,Cheung:2011wn}
in the calculations of the $125.5$-GeV Higgs-mediated contributions to
the EDMs. 
We start by giving the relevant interaction Lagrangian as
\begin{eqnarray}
{\cal L} & = &
  -\; \frac{i}{2}\,d^E_f\,F^{\mu\nu}\,\bar{f}\,\sigma_{\mu\nu}\gamma_5\,f\
-\ \frac{i}{2}\,d_q^C\,G^{a\,\mu\nu}\,\bar{q}\,\sigma_{\mu\nu}\gamma_5 T^a\,q
\nonumber \\
 &&
+\frac{1}{3}\,d^{\,G}\,f_{abc}\,G^a_{\rho\mu}\,\widetilde{G}^{b\,\mu\nu}\,
{G^c}_{\nu}^{~~\rho}\ + \
\sum_{f,f'}     C_{ff'}     (\bar{f}f)
(\bar{f'}i\gamma_5f')\; ,
\end{eqnarray}
where
$F^{\mu\nu}$ and  $G^{a\,\mu\nu}$ are  the  electromagnetic and
strong  field strengths, respectively,  the $T^a=\lambda^a/2$  are the
generators    of    the     SU(3)$_C$    group and
$\widetilde{G}^{\mu\nu} = \frac{1}{2} \epsilon^{\mu\nu\lambda\sigma}
G_{\lambda\sigma}$ is the dual  of the SU(3)$_c$ field-strength tensor
$G_{\lambda\sigma}$.

We denote the  EDM of a fermion  by $d^E_f$ and the chromoelectric dipole moment
(CEDM)  of a quark by $d^C_q$.
The major Higgs-mediated contribution comes from 
the two-loop Barr--Zee-type diagrams, labeled as 
\begin{eqnarray}
\left(d_f^E\right)^H  =  (d^E_f)^{\rm BZ}\,; \ \ \
\left(d_q^C\right)^H  =  (d^C_q)^{\rm BZ}\,,
\end{eqnarray}
the details of which will be discussed below.
For the Weinberg operator, we consider the contributions from
the Higgs-mediated two-loop diagrams:
\begin{equation}
\label{eq:dg}
(d^{\,G})^{H}\ =\ \frac{4\sqrt{2}\, G_F\, g_s^3}{(4\pi)^4}
\sum_{q=t,b} \, g^S_{H\bar{q}q}\,
  g^P_{H\bar{q}q}\,h(z_{Hq})\,,
\end{equation}
where $z_{Hq} \equiv M_H^2/m_q^2$ with $M_H=125.5$ GeV
and, for the loop function $h(z_{Hq})$,
we refer to Ref.~\cite{Dicus:1989va}.
We note, in passing, that $(d^{\,G})^{H}$ depends on the $H$ couplings to
the third-generation quarks only.
For the four-fermion operators, we consider
the $t$-channel exchanges of the CP-mixed state $H$,
which give rise to the CP-odd coefficients as follows~\cite{Ellis:2008zy}:
\begin{equation}
  \label{eq:cff}
(C_{ff'})^H\ =\ g_f\, g_{f'}\,
\frac{g^S_{H\bar{f}f}\,g^P_{H\bar{f'}f'}}{M_{H}^2}\; .
\end{equation}

\subsection{Two-loop Barr--Zee EDMs}
We consider both the Barr--Zee diagrams
mediated by the $\gamma$-$\gamma$-$H$ couplings~\cite{Ellis:2008zy} and
by the $\gamma$-$H$-$Z$ couplings~\cite{Giudice:2005rz,Li:2008kz}.
More explicitly the contributions from the two-loop
Higgs-mediated Barr--Zee-type diagrams can be decomposed into
two parts:
\begin{equation}
\left(d^E_f\right)^{\rm BZ} =
\left(d^E_f\right)^{\gamma H}+ \
\left(d^E_f\right)^{Z H}
\end{equation}
where
\begin{eqnarray}
\label{eq:def}
(-Q_f)^{-1}\times \left(\frac{d_f^E}{e}\right)^{\gamma H}
\!&=&\! \sum_{q=t,b}\Bigg\{
\frac{3\alpha_{\rm em}^2\,Q_q^2\,m_f}{8\pi^2s_W^2M_W^2}
\left[
g^P_{H\bar{f}f} g^S_{H\bar{q}q}\,f(\tau_{qH})
+g^S_{H\bar{f}f} g^P_{H\bar{q}q}\,g(\tau_{qH})
\right]\Bigg\}  \nonumber \\
&&\! +\
\frac{\alpha_{\rm em}^2\,m_f}{8\pi^2s_W^2M_W^2}
\left[
g^P_{H\bar{f}f} g^S_{H\tau^+\tau^-}\,f(\tau_{\tau H})
+g^S_{H\bar{f}f} g^P_{H\tau^+\tau^-}\,g(\tau_{\tau H})
\right] \nonumber \\
&&\! -\frac{\alpha_{\rm em}^2\,m_f}{32\pi^2s_W^2M_W^2}\,
g^P_{H\bar{f}f}g_{HVV}\,{\cal J}_W^\gamma(M_{H})
\end{eqnarray}
with $\tau_{xH} =  m_x^2/M_{H}^2$.
For the loop functions $f(\tau)$ and $g(\tau)$
we refer to, for example, Refs.~\cite{Ellis:2008zy,Ellis:2010xm}
and references therein.
The loop function ${\cal J}_W^{G=\gamma,Z}(M_{H})$
for the $W$-loop contributions is given by~\cite{Abe:2013qla}
\begin{eqnarray}
&&\hspace{-0.5cm}
{\cal J}^G_W(M_H)=\frac{2M_W^2}{M_H^2-M_G^2}\Bigg\{
-\frac{1}{4}\left[
\left(6-\frac{M_G^2}{M_W^2}\right)+
\left(1-\frac{M_G^2}{2M_W^2}\right)\frac{M_H^2}{M_W^2}
\right]\,\left[I_1(M_W,M_H)-I_1(M_W,M_G)\right]
\nonumber \\[5mm] && \hspace{-0.0cm} +
\left[\left(-4+\frac{M_G^2}{M_W^2}\right)+\frac{1}{4}
\left(6-\frac{M_G^2}{M_W^2}\right)+\frac{1}{4}
\left(1-\frac{M_G^2}{2M_W^2}\right)\frac{M_H^2}{M_W^2}
\right]\,\left[I_2(M_W,M_H)-I_2(M_W,M_G)\right]\Bigg\}
\nonumber \\
\end{eqnarray}
where
\begin{equation}
I_{1}(m_1,m_2)=-2\frac{m_2^2}{m_1^2} f\left(\frac{m_1^2}{m_2^2}\right)\,,
\ \ \
I_{2}(m_1,m_2)=-2\frac{m_2^2}{m_1^2} g\left(\frac{m_1^2}{m_2^2}\right)\,.
\end{equation}
We note that, for large $\tau$, $f(\tau)\sim 13/18+(\ln\tau)/3 $ and
$g(\tau)\sim 1+(\ln\tau)/2$~\cite{Barr:1990vd}.
Also, $(d_f^E)^{Z H}$ is given by
\begin{eqnarray}
\left(\frac{d_f^E}{e}\right)^{Z H} &=&
\frac{\alpha_{\rm em}^2 v_{Z\bar{f}f}}{16\sqrt{2}\pi^2c_W^2s_W^4}\,
\frac{m_f}{M_W}\,
\sum_{q=t,b}
\frac{3Q_qm_q}{\sqrt{2}M_W}\,
\nonumber \\
&\times & \Bigg[\
g^S_{H\bar{f}f} \left(v_{Z\bar{q}q} g^P_{H\bar{q}q}\right)\,
\frac{m_q}{M_H^2}\,
\int_0^1 {\rm d}x\frac{1}{x}
J\left(r_{ZH},\frac{r_{qH}}{x(1-x)}\right)
\nonumber \\
&& \hspace{0.6cm}+
g^P_{H\bar{f}f} \left(v_{Z\bar{q}q} g^S_{H\bar{q}q}\right)\,
\frac{m_q}{M_H^2}\,
\int_0^1 {\rm d}x\frac{1-x}{x}
J\left(r_{ZH},\frac{r_{qH}}{x(1-x)}\right) \Bigg]
\nonumber \\
&-& \frac{\alpha_{\rm em}^2 v_{Z\bar{f}f}}{16\sqrt{2}\pi^2c_W^2s_W^4}\,
\frac{m_f}{M_W}\,\frac{m_\tau}{\sqrt{2}M_W}\,
\nonumber \\
&\times &  \Bigg[\
g^S_{H\bar{f}f} \left(v_{Z\tau^+\tau^-} g^P_{H\tau^+\tau^-}\right)\,
\frac{m_\tau}{M_H^2}\,
\int_0^1 {\rm d}x\frac{1}{x}
J\left(r_{ZH},\frac{r_{\tau H}}{x(1-x)}\right)
\nonumber \\
&& \hspace{0.6cm}+
g^P_{H\bar{f}f} \left(v_{Z\tau^+\tau^-} g^S_{H\tau^+\tau^-}\right)\,
\frac{m_\tau}{M_H^2}\,
\int_0^1 {\rm d}x\frac{1-x}{x}
J\left(r_{ZH},\frac{r_{\tau H}}{x(1-x)}\right) \Bigg]
\nonumber \\
&&\! +\frac{\alpha_{\rm em}^2\,v_{Z\bar{f}f}\,m_f}{32\pi^2s_W^4M_W^2}\,
g^P_{H\bar{f}f}g_{HVV}\,{\cal J}_W^Z(M_{H})\,,
\end{eqnarray}
with $r_{xy} \equiv M_x^2/M_y^2$. For the loop function
$J(a,b)$ we again refer to, for example, Refs.~\cite{Ellis:2008zy,Ellis:2010xm}
and references therein.
The $Z$-boson couplings to the quarks and leptons are given by
\begin{eqnarray}
{\cal L}_{Z\bar{f}f} = -\,g_Z\,
\bar{f}\,\gamma^\mu\,\left(v_{Z\bar{f}f} - a_{Z\bar{f}f} \gamma_5\right)\,f\,Z_\mu
\end{eqnarray}
with $v_{Z\bar{f}f}=T_{3L}^f/2-Q_f s_W^2$ and $a_{Z\bar{f}f}=T_{3L}^f/2$ and
$g_Z=g/c_W=(e/s_W)/c_W$.
For the SM quarks and leptons,
$T_{3L}^{\,u, \nu}=+1/2$ and $T_{3L}^{\,d, e}=-1/2$.

In addition to EDMs, the two-loop  Higgs-mediated
Barr-Zee graphs also generate CEDMs
of the light quarks $q_l=u,d$, which take the form:
\begin{eqnarray}
\label{eq:cedm}
\left(d_{q_l}^C\right)^{\rm BZ} \!&=&\! -
\frac{g_s\,\alpha_s\,\alpha_{\rm em}\,m_{q_l}}{16\pi^2s_W^2M_W^2}
\sum_{q=t,b}\left[
g^P_{H\bar{q}_lq_l} g^S_{H\bar{q}q}\,f(\tau_{qH})
+g^S_{H\bar{q}_lq_l} g^P_{H\bar{q}q}\,g(\tau_{qH})
\right]\,.  
\end{eqnarray}

\subsection{Observable EDMs}
In this subsection, we briefly review the dependence of the
Thallium, neutron, Mercury, deuteron, Radium, and thorium-monoxide EDMs on the
EDMs and/or CEDMs of quarks and leptons, and on 
the coefficients of  the dimension-six 
Weinberg operator and the four-fermion operators.

\subsubsection{Thallium EDM}
The Thallium EDM receives   contributions   mainly   from   two
terms~\cite{KL,Pospelov:2005pr}:
\begin{equation}
\label{eq:dTl}
d_{\rm Tl}\,[e\,{\rm cm}]\ =\ -585\cdot d_e^E\,[e\,{\rm cm}]\:
-\: 8.5\times 10^{-19}\,[e\,{\rm cm}]\cdot (C_S\,{\rm TeV}^2)\: +\ \cdots\,,
\end{equation}
where $d^E_e$ is the electron EDM and $C_S$ is the coefficient of
the CP-odd electron-nucleon interaction
${\cal L}_{C_S}=C_S\,\bar{e}i\gamma_5\,e \bar{N}N$, which is given by
\begin{equation}
C_S = C_{de}\frac{29\,{\rm MeV}}{m_d}
+ C_{se}\frac{\kappa\times 220\,{\rm MeV}}{m_s}
+ (0.1\,{\rm GeV})\,
\frac{m_e}{v^2}
\frac{g^S_{H_igg}g^P_{H\bar{e}e}}{M_{H}^2}
\end{equation}
with $\kappa\equiv\langle N |
m_s \bar{s} s | N \rangle/220~{\rm MeV} \simeq 0.50\pm0.25$
and
\begin{equation}
g^S_{H_igg}=\sum_{q=t,b}\left\{\frac{2\,x_q}{3}g_{H_i\bar{q}q}^S
\right\}\,,
\end{equation}
with $x_t=1$ and $x_b=1-0.25 \kappa$.

\subsubsection{Thorium-Monoxide EDM}
Similar to the Thallium EDM, 
the thorium-monoxide EDM is given by~\cite{dfh:2011}:
\begin{equation}
\label{eq:dThO}
d_{\rm ThO}\,[e\,{\rm cm}]\ =\ {\cal F}_{\rm ThO}\left\{d_e^E\,[e\,{\rm cm}]\:
+\: 1.6\times 10^{-21}\,[e\,{\rm cm}]\, (C_S\,{\rm TeV}^2)\right\}\: +\ \cdots\,.
\end{equation}
Currently, the experimental constraint is given on the quantity
$\left|d_{\rm ThO}/{\cal F}_{\rm ThO}\right|$.

\subsubsection{Neutron EDM}
For the neutron EDM we take the hadronic approach with
the QCD sum-rule technique.
In this approach, the neutron EDM is given
by~\cite{qcdsumrule1,qcdsumrule2,Demir:2002gg,Demir:2003js,Olive:2005ru}
\begin{eqnarray}
  \label{eq:dnQCD}
d_n \!&=&\! d_n(d_q^E\,,d_q^C)\: +\: d_n(d^{\,G})\: +\: d_n(C_{bd})\:
+\: \cdots\,, \nonumber \\
d_n(d_q^E\,,d_q^C) \!&=&\! (1.4\pm 0.6)\,(d_d^E-0.25\,d_u^E)\:
+\: (1.1\pm 0.5)\,e\,(d_d^C+0.5\,d_u^C)/g_s\,,
\nonumber \\
d_n(d^{\,G}) \!&\sim&\! \pm\, e\, (20\pm 10)~{\rm MeV} \,d^{\,G}\,,
\nonumber \\
d_n (C_{bd}) \!&\sim &\! \pm\, e\, 2.6\times 10^{-3}~{\rm GeV}^2\,
\left[\frac{C_{bd}}{m_b}\: +\: 0.75\frac{C_{db}}{m_b}\right]\; ,
\end{eqnarray}
where  $d_q^E$ and $d_q^C$  should be  evaluated at the electroweak (EW)
scale and
$d^{\,G}$  at  the 1  GeV  scale, for which 
$d^G\big|_{1~{\rm  GeV}} \simeq (\eta^G/0.4)\, d^G\big|_{\rm
EW}  \simeq 8.5\,  d^G\big|_{\rm EW}$~\cite{Demir:2002gg}
taking $\eta^G=3.4$~\cite{Arnowitt:1990eh,Ibrahim:1997gj}.
In the numerical estimates
we take the positive sign for both $d_n(d^{\,G})$ and $d_n (C_{bd})$.

\subsubsection{Mercury EDM}
Using the QCD sum rules~\cite{Demir:2003js,Olive:2005ru}, we estimate
the Mercury EDM as
\begin{eqnarray}
d^{\rm \,I\,,II\,,III\,,IV}_{\rm Hg} \!&=&\!
d^{\rm \,I\,,II\,,III\,,IV}_{\rm Hg}[S]
+10^{-2} d_e^E 
+(3.5\times 10^{-3} {\rm GeV})\,e\,C_S
\nonumber \\ \!&&\!
+\ (4\times 10^{-4}~{\rm GeV})\,e\,\left[C_P+
\left(\frac{Z-N}{A}\right)_{\rm Hg}\,C^\prime_P\right]\,,
\end{eqnarray}
where $d^{\rm \,I\,,II\,,III\,,IV}_{\rm Hg}[S]$
denotes the Mercury EDM induced by the Schiff moment.
The parameters $C_P$ and $C^\prime_P$  are the couplings of
electron-nucleon interactions as in
${\cal L}_{C_P}\ =\ C_P\,\bar{e}e\,\bar{N}i\gamma_5 N\: +\:
C^\prime_P\,\bar{e}e\,\bar{N}i\gamma_5 \tau_3  N$ and they are given by
\cite{Ellis:2008zy}
\begin{eqnarray}
C_P & \simeq &
-\,375~{\rm MeV}\,\sum_{q=c,s,t,b} \frac{C_{eq}}{m_q}\,,\nonumber \\
C^\prime_P & \simeq\ & -\,806~{\rm MeV}\,\frac{C_{ed}}{m_d}\,
-\,181~{\rm MeV}\,\sum_{q=c,s,t,b} \frac{C_{eq}}{m_q}\,.
\end{eqnarray}
In this work, we take $d^{\rm \,I}_{\rm Hg}[S]$
for the Schiff-moment induced Mercury EDM, which is given by~\cite{Ellis:2011hp}
\begin{eqnarray}
  \label{eq:dHgI}
d^{\rm \,I}_{\rm Hg}[S]\ & \simeq\ &
1.8 \times 10^{-3}\, e\,\bar{g}^{(1)}_{\pi NN}\,/{\rm GeV}\,,
\end{eqnarray}
where
\begin{eqnarray}
\bar{g}^{(1)}_{\pi NN} \!&=&\!
2^{+4}_{-1}\times 10^{-12}\,\frac{(d_u^C-d_d^C)/g_s}{10^{-26}{\rm cm}}\,
\frac{|\langle \bar{q} q\rangle |}{(225\,{\rm MeV})^3} \nonumber \\
&& -\, 8\times 10^{-3} {\rm GeV}^3\,
\left[\frac{0.5C_{dd}}{m_d}+3.3\kappa\frac{C_{sd}}{m_s}
+(1-0.25\kappa)\frac{C_{bd}}{m_b} \right]\,.
\end{eqnarray}

\subsubsection{Deuteron EDM}
For the deuteron EDM, we use~\cite{Lebedev:2004va,Ellis:2008zy}:
\begin{eqnarray}
d_{D} \!&\simeq &\! -\left[5^{+11}_{-3} + (0.6\pm 0.3)
  \right]\,e\,(d^C_u-d^C_d)/g_s
\nonumber \\
\!&&\! -(0.2\pm 0.1)\,e\,(d^C_u+d^C_d)/g_s\:
+\: (0.5 \pm 0.3) (d^E_u+d^E_d) \nonumber \\
\!&&\! +(1\pm 0.2)\times 10^{-2}\,e\,{\rm GeV}^2\,
\left[\frac{0.5C_{dd}}{m_d}+3.3\kappa\frac{C_{sd}}{m_s}
+(1-0.25\kappa)\frac{C_{bd}}{m_b} \right] \nonumber \\
&& \pm\ e\, (20\pm 10)~{\rm MeV} \,d^{\,G}\; .
\end{eqnarray}
In  the above,  $d^{\,G}$  is evaluated  at the  $1$ GeV scale,
and the coupling coefficients $g_{d,s,b}$ appearing in $C_{dd,sd,bd}$
are computed at energies 1~GeV, 1~GeV and $m_b$, respectively.
All other  EDM  operators  are calculated  at  the EW  scale.
In the numerical estimates
we take the positive sign for $d^{\,G}$.

\subsubsection{Radium EDM}
For the EDM of $^{225}$Ra,
we use~\cite{Ellis:2011hp}:
\begin{eqnarray}
d_{\rm Ra} \simeq
d_{\rm Ra}[S] \simeq
-8.7 \times 10^{-2}\, e\,\bar{g}^{(0)}_{\pi NN}\,/{\rm GeV}
+3.5 \times 10^{-1}\, e\,\bar{g}^{(1)}_{\pi NN}\,/{\rm GeV}\,,
\end{eqnarray}
where
\begin{equation}
\bar{g}^{(0)}_{\pi NN} \!=\!
0.4\times 10^{-12}\,\frac{(d_u^C+d_d^C)/g_s}{10^{-26}{\rm cm}}\,
\frac{|\langle \bar{q} q\rangle |}{(225\,{\rm MeV})^3}\,.
\end{equation}
We  note that  the $\bar{g}^{(1)}_{\pi  NN}$ contribution  to
the  Radium EDM is about  200 times larger than that to the
Mercury EDM $d^{\rm \,I}_{\rm Hg}[S]$~\cite{Engel:2003rz}.

\section{Numerical Analysis}

The non-observation of EDMs for Thallium~\cite{Regan:2002ta},
neutron~\cite{Baker:2006ts},
Mercury~\cite{Griffith:2009zz},
and thorium monoxide~\cite{Baron:2013eja}
constrains the CP-violating phases through
\begin{eqnarray}
&&
|d_{\rm Tl}| \leq d_{\rm Tl}^{\rm EXP}\,, \ \
|d_{\rm n}| \leq d_{\rm n}^{\rm EXP}\,, \ \ \nonumber \\[3mm]
&&
|d_{\rm Hg}| \leq d_{\rm Hg}^{\rm EXP}\,, \ \
|d_{\rm ThO}/{\cal F}_{\rm ThO}| \leq d_{\rm ThO}^{\rm EXP}\,,
\end{eqnarray}
with the current experimental bounds
\begin{eqnarray}
\label{eq:exp}
&&
d_{\rm Tl}^{\rm EXP}=9\times 10^{-25}\, e\,{\rm cm}\,, \ \
d_{\rm n}^{\rm EXP}=2.9\times 10^{-26}\, e\,{\rm cm}\,, \ \ \nonumber \\[3mm]
&&
d_{\rm Hg}^{\rm EXP}=3.1\times 10^{-29}\, e\,{\rm cm}\,, \ \
d_{\rm ThO}^{\rm EXP}=8.7\times 10^{-29}\, e\,{\rm cm}\,. \ \
\end{eqnarray}
For the normalization of the deuteron and Radium EDMs, we have taken
the projected experimental sensitivity~\cite{Semertzidis:2003iq} to be
$d_{\rm D}^{\rm PRJ}=3\times 10^{-27}\, e\,{\rm cm}$ and
$d_{\rm Ra}^{\rm PRJ}=1\times 10^{-27}\, e\,{\rm cm}$, respectively.
The chosen value for $d_{\rm Ra}^{\rm PRJ}$
is near to a sensitivity which can be achieved in
one day of  data-taking~\cite{Willmann}.
On the other hand,
the future Higgs-boson data may shrink the CL regions that we obtained in 
Fig.~\ref{fig:hfit}.
Nevertheless, we have to emphasize that the combined constraint 
on $|C_u^P|$ from all the current EDM measurements is at the level of
$10^{-2}$ at 95\% CL without any further assumptions beyond the 125.5 GeV
Higgs-mediated contributions, see Eq.~(\ref{eq:cup-limit}).
The future Higgs-boson data alone cannot further reduce
such a strong constraint on $|C_u^P|$ while 
the deuteron and Radium EDMs are capable of probing 
$|C_u^P|\lsim 10^{-2}$ with the estimates of the projected sensitivities.

\subsection{(C)EDMs of quarks and leptons and $d^G$}
In this subsection, we analyze the contributions of 
the Higgs boson $H$ with the mass $125.5$ GeV in the 2HDM framework to
\begin{itemize}
\item EDMs of electron and up and down quarks:
$d^E_f=(d^E_f)^{\rm BZ}=(d^E_f)^{\gamma H}+(d^E_f)^{ZH}$ with $f=e,u,d$,
\item CEDMs of up and down quarks: $d^C_q=(d^C_q)^{\rm BZ}$ with $q=u,d$, and
\item Coefficient of the Weinberg operator $d^G$,
\end{itemize}
together with their constituent contributions, taking the benchmark point
\begin{equation}
C_u^S=C_u^P=1/2\,, 
\end{equation}
while varying $C_v$.

\begin{figure}[t!]
\hspace{ 0.0cm}
\vspace{-0.5cm}
\centerline{\epsfig{figure=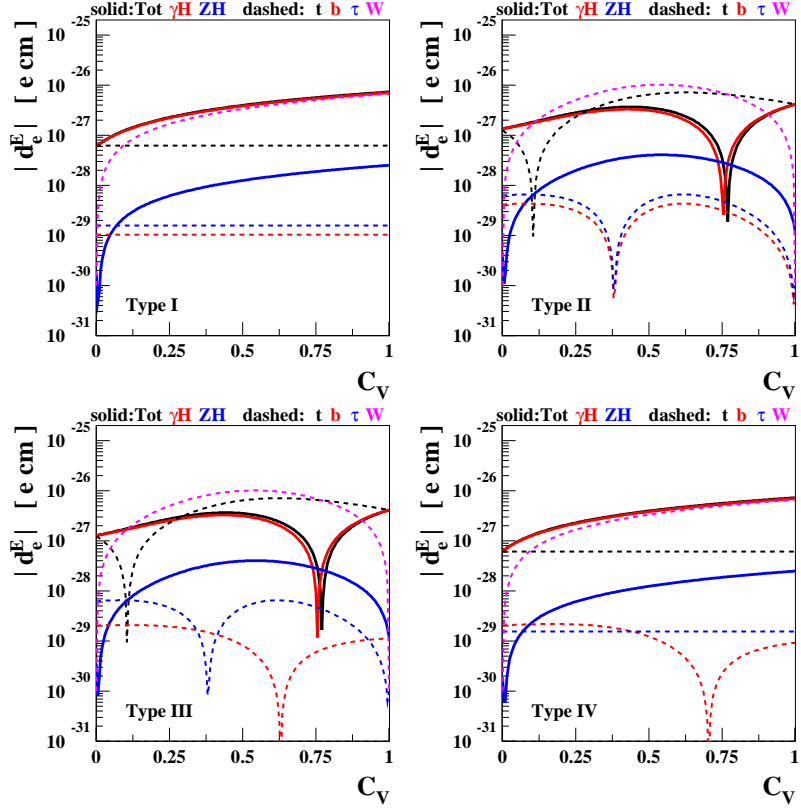,height=12cm,width=12cm}}
\vspace{-0.5cm}
\caption{\it The absolute values of the electron EDM as functions of $C_v$
in units of $e\,{\rm cm}$
when $C_u^S=C_u^P=1/2$ for the types I -- IV of 2HDMs. The red and blue solid
lines are for $(d^E_e)^{\gamma H}$ and $(d^E_e)^{ZH}$, respectively, and the black
solid lines are for the total sum. The constituent contributions from top, bottom, 
tau, and W-boson loops are denoted by the 
dashed black, red, blue, and magenta lines, respectively.
}
\label{fig:dee}
\end{figure}
In Fig.~\ref{fig:dee}, we show the electron EDM as a function of $C_v$
in units of $e\,{\rm cm}$
with $C_u^S=C_u^P=1/2$ for the types I -- IV of 2HDMs. The red and blue solid
lines are for $(d^E_e)^{\gamma H}$ and $(d^E_e)^{ZH}$, respectively, and the black
solid lines are for the total sum. The constituent contributions from 
the top, bottom, tau, and W-boson loops are denoted by the
dashed black, red, blue, and magenta lines, respectively. 
In all types of 2HDMs,
we observe that $(d^E_e)^{\gamma H}$, the contribution from 
the $\gamma$-$H$ Barr-Zee diagram, dominates over $(d^E_e)^{Z H}$,
which is suppressed by the factor $v_{Z\bar e e}=-1/4+s_W^2$. 
Also, the $W$-boson loop contribution is dominant
in types I and IV when $C_v\gsim 0.1$, and the top and $W$-boson 
loop contributions
are comparable in Types II and III.

Keeping only the top and $W$-loop contributions in the $\gamma$-$H$ Barr-Zee
diagram and neglecting the $Z$-$H$ Barr-Zee diagram,  
the electron EDM satisfies
\begin{eqnarray}
\left(\frac{d_e^E}{e}\right)_{\rm I,IV} &\propto &
\left\{\frac{16}{3}\left[-f(\tau_{tH})+g(\tau_{tH})\right]\, C_u^S
+ C_v\,{\cal J}_W^\gamma(M_{H})\right\} \,C_u^P\,, 
\\[2mm]
\left(\frac{d_e^E}{e}\right)_{\rm II,III} &\propto &
\left\{\frac{16}{3}\left[
 t_\beta^2 C_u^S\,f(\tau_{tH})
+\frac{O_{\phi_1 i}}{c_\beta} \,g(\tau_{tH})\right] 
-t_\beta^2 C_v\,{\cal J}_W^\gamma(M_{H})\right\}\,C_u^P \nonumber \;,
\end{eqnarray}
for Type I,IV and II, III, respectively: see Eq.~(\ref{eq:def}).
Numerically, $f(\tau_{tH})\simeq 0.98$, $g(\tau_{tH})\simeq 1.4$, and ${\cal
J}_W^\gamma(M_{H})\simeq 12$.
%
\begin{figure}[t!]
\hspace{ 0.0cm}
\vspace{-0.5cm}
\centerline{\epsfig{figure=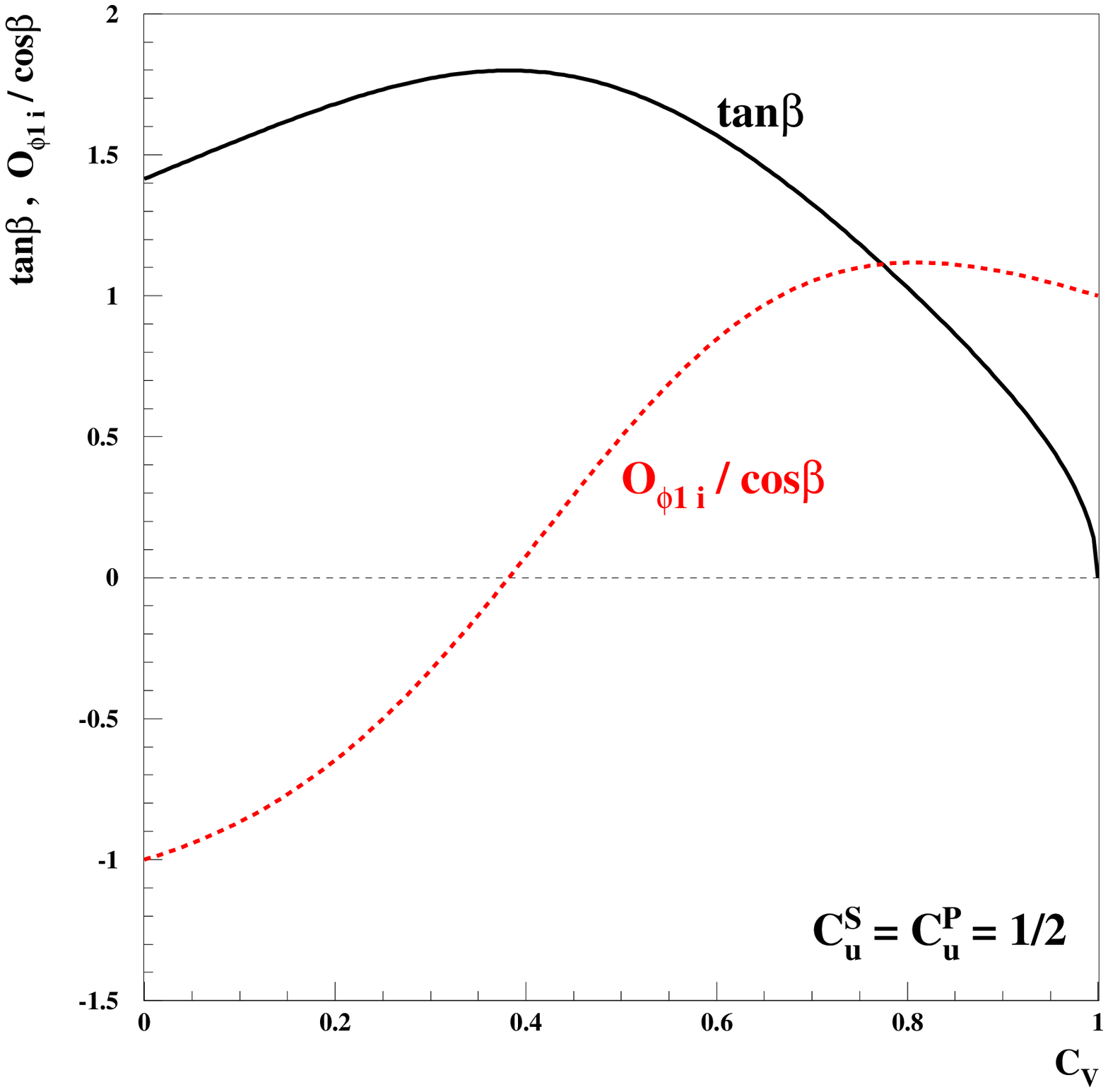,height=8cm,width=8cm}}
\vspace{-0.5cm}
\caption{\it $\tan\beta$ and $O_{\phi_1i}/\cos\beta$ as functions of $C_v$
taking $C_u^S=C_u^P=1/2$.
}
\label{fig:coupl}
\end{figure}
We observe that the electron EDM is
overall proportional to $C_u^P$ and it flips the sign according to 
the change in the sign of $C_u^P$.
The top and $W$ contributions have the same signs, and 
the top-quark contributions
are independent of $C_v$ in Types I and IV. 
Also, note that the two top-quark contributions in Types I and IV
cancel each other 
so that the top-quark contribution is suppressed compared to 
that in Types II and III.
For the reference point $C_u^S=C_u^P=1/2$, we show $\tan\beta$ and 
$O_{\phi_1i}/\cos\beta$ as functions of $C_v$ in Fig.~\ref{fig:coupl}
\footnote{
Note that $\sin\beta=0$ when $C_v=1$  for non-zero $C_u^P$ independent
of $C_u^S$, see Eq.~(\ref{eq:sbsq}).}.
When $C_v\gsim 0.4$, $O_{\phi_i1}$ is positive and we see that
the top and $W$ contributions have the opposite signs
in Types II and III, which
leads to a large cancellation 
between the top (dashed black lines) and $W$ (dashed magenta lines)
contributions around $C_v=0.75$
in Types II and III: see the upper-right and lower-left frames of
Fig.~\ref{fig:dee}.
Since $O_{\phi_i1}<0$ when $C_v\lsim 0.4$, the two top-quark contributions
in Types II and III cancel each other and thus explains the dips 
in the constituent contributions from top loops (black dashed lines)
around $C_v=0.1$ in Types II and III.

\begin{figure}[t!]
\hspace{ 0.0cm}
\vspace{-0.5cm}
\centerline{\epsfig{figure=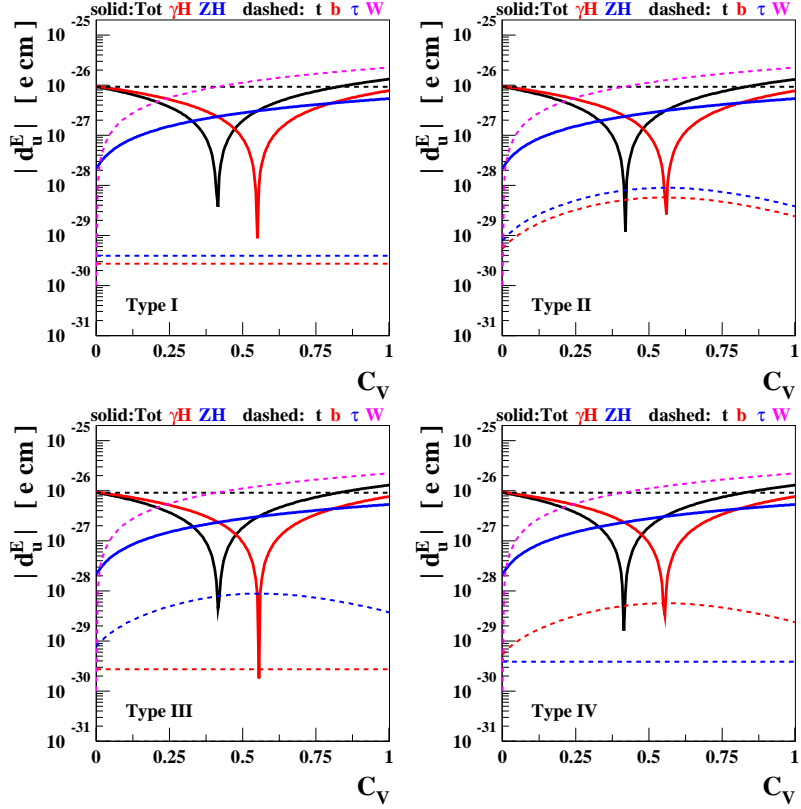,height=12cm,width=12cm}}
\vspace{-0.5cm}
\caption{\it The same as in Fig.~\ref{fig:dee} but for
the up-quark EDM.
}
\label{fig:deu}
\end{figure}

In Fig.~\ref{fig:deu}, we show the absolute values of the
up-quark EDM as a function of $C_v$ in units of $e\,{\rm cm}$
with $C_u^S=C_u^P=1/2$ for Types I -- IV of 2HDMs. The labeling of the lines
is the same as in Fig.~\ref{fig:dee}. We find that
the contributions from the $Z$-$H$ Barr-Zee (solid blue lines) diagrams
are comparable to those from the
$\gamma$-$H$ Barr-Zee (solid red lines) ones, and
the $Z$-$H$ Barr-Zee contributions are dominated by the $W$-boson loops.
In this case, similar to the electron EDM case, the up-quark EDM satisfies
\begin{eqnarray}
\left(\frac{d_u^E}{e}\right)_{\rm I,II,III,IV} &\propto &
\left\{\left[\frac{16}{3}\left(f(\tau_{tH})+g(\tau_{tH})\right)\, C_u^S
- C_v\,{\cal J}_W^\gamma(M_{H})\right]\times\left(-\frac{2}{3}\right)
+\frac{v_{Z\bar u u}}{s_W^2}\,C_v\,{\cal J}_W^Z(M_{H})\right\} \,C_u^P
\nonumber \\
\end{eqnarray}
which are independent of the 2HDM type. We find ${\cal J}_W^Z(M_{H})\simeq 5.5$.
The top-quark contribution is negative and
the $W$-loop contribution is positive because $v_{Z\bar u u}>0$. One may see 
$(d^E_u)^{\gamma H}$ vanishes when the first two terms cancel and
a cancellation may also occur between
$(d^E_u)^{\gamma H}$ and $(d^E_u)^{Z H}$.
The former cancellation explains the dips of $|(d^E_u)^{\gamma H}|$ (red solid lines)
around $C_v=0.55$ and the latter one explains the dips of the 
total (black solid lines) up-quark EDMs around $C_v=0.4$
%
%

\begin{figure}[t!]
\hspace{ 0.0cm}
\vspace{-0.5cm}
\centerline{\epsfig{figure=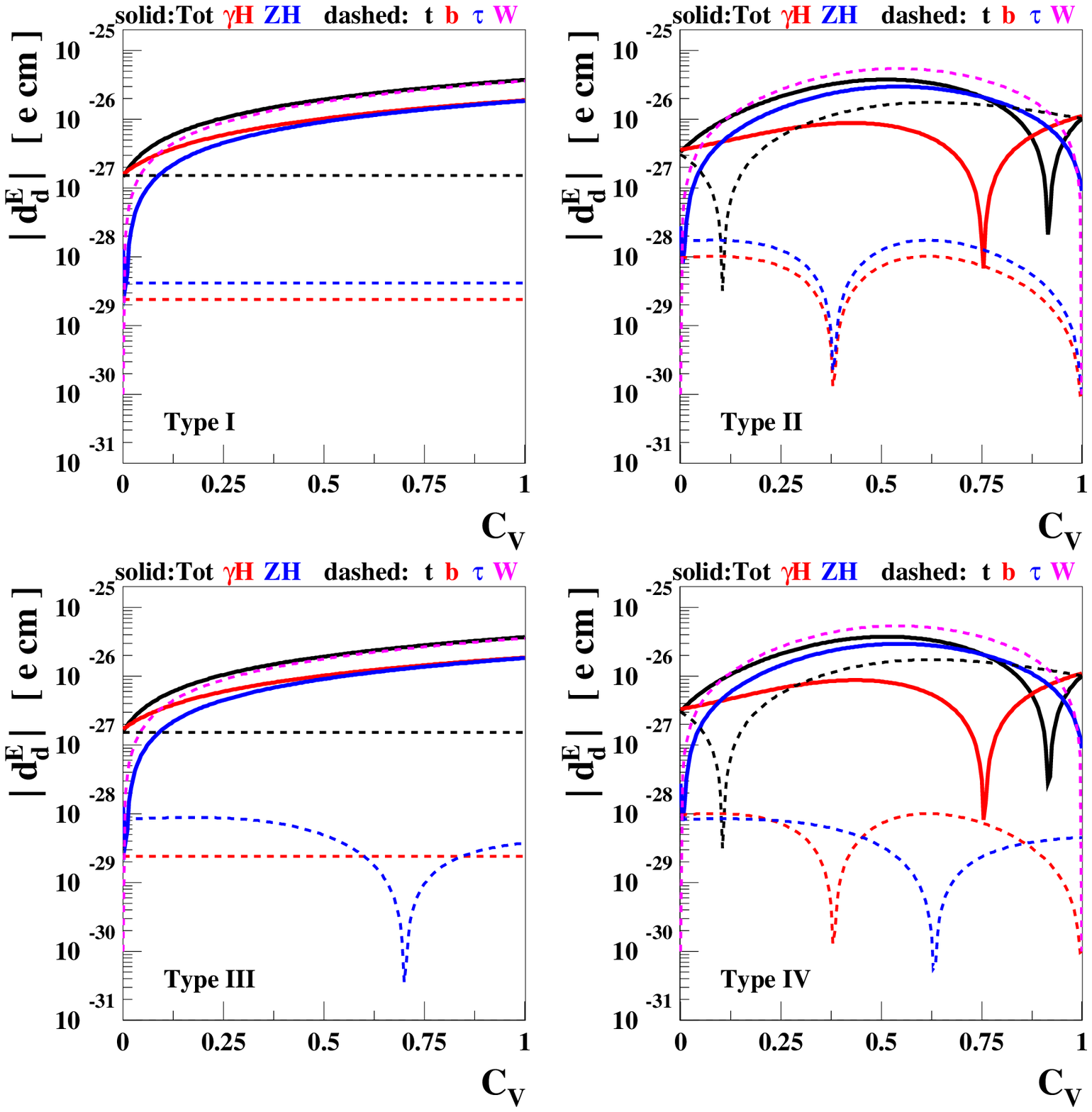,height=12cm,width=12cm}}
\vspace{-0.5cm}
\caption{\it The same as in Fig.~\ref{fig:dee} but for 
the down-quark EDM.
}
\label{fig:ded}
\end{figure}
In Fig.~\ref{fig:ded}, we show the absolute values of the
down-quark EDM as a function of $C_v$ in units of $e\,{\rm cm}$
with $C_u^S=C_u^P=1/2$ for the Types I -- IV of 2HDMs. The labeling of lines
is the same as in Fig.~\ref{fig:dee}.  
Similar to  the up-quark EDM, the $\gamma$-$H$
Barr--Zee diagram is dominated by
the top and $W$ loops and the $Z$-$H$ one by the $W$ loop.
Considering these three dominant constituent contributions, 
the down-quark EDM satisfies
\begin{eqnarray}
\left(\frac{d_d^E}{e}\right)_{\rm I,III} &\propto &
\left\{\left[\frac{16}{3}\left(-f(\tau_{tH})+g(\tau_{tH})\right)\, C_u^S
+ C_v\,{\cal J}_W^\gamma(M_{H})\right]\times\left(\frac{1}{3} \right)
-\frac{v_{Z\bar d d}}{s_W^2}\,C_v\,{\cal J}_W^Z(M_{H})\right\} \,C_u^P\,,
\nonumber \\[2mm]
\left(\frac{d_d^E}{e}\right)_{\rm II,IV} &\propto &
\left\{\left[\frac{16}{3}\left(
 t_\beta^2 C_u^S\,f(\tau_{tH})
+\frac{O_{\phi_1 i}}{c_\beta} \,g(\tau_{tH})\right) 
-t_\beta^2 C_v\,{\cal J}_W^\gamma(M_{H})\right]\times\left(\frac{1}{3} \right)
+\frac{v_{Z\bar d d}}{s_W^2}\,t_\beta^2 C_v\,{\cal J}_W^Z(M_{H})\right\} \,C_u^P\,.
\nonumber \\
\end{eqnarray}
First we note that all three contributions in Types I and III are 
positive because $v_{Z\bar d d}<0$. As in the electron EDM, we find the 
top-quark contributions are
independent of $C_v$.
In Types II and IV, the two top-quark contributions cancel each other 
around $C_v=0.1$ (dips of the black dashed lines)
and they turn to be positive when $C_v\gsim 0.1$.
Since both of the $W$ loop contributions are negative,
the cancellation between the 
positive top and negative $W$ contributions
explains the dips of $|(d^E_d)^{\gamma H}|$ around $C_v=0.75$ (solid red lines)
and those of the total sum (black solid lines) around 
$C_v=0.9$. Note $t_\beta^2C_v$
decreases as $C_v$ increases when $C_v\gsim 0.5$.

\begin{figure}[t!]
\hspace{ 0.0cm}
\vspace{-0.5cm}
\centerline{\epsfig{figure=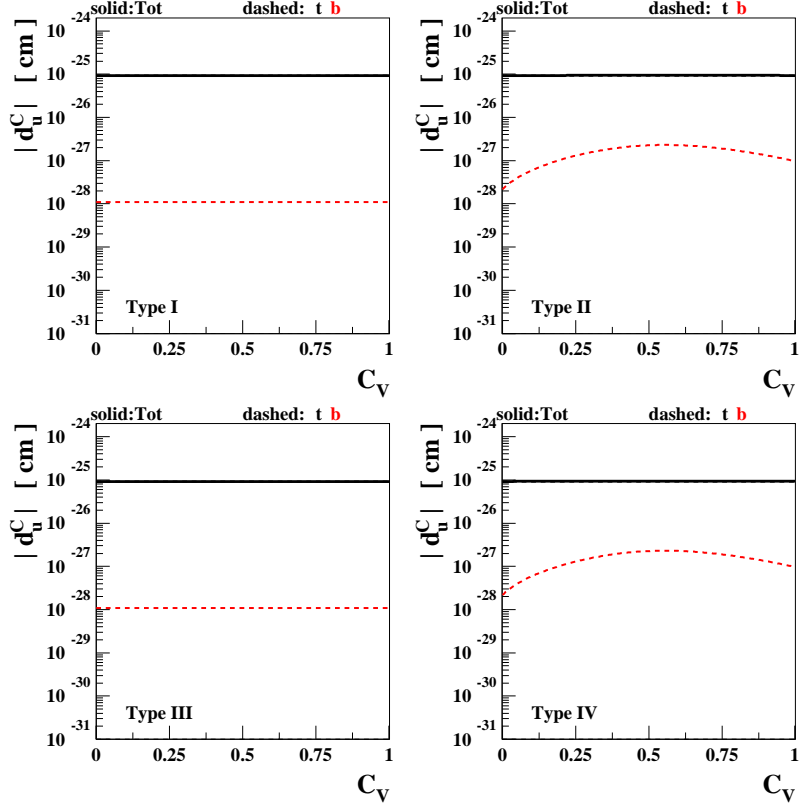,height=12cm,width=12cm}}
\vspace{-0.5cm}
\caption{\it The absolute values of the up-quark CEDM as functions of $C_v$
in units of ${\rm cm}$
when $C_u^S=C_u^P=1/2$ for the types I -- IV of 2HDMs. 
The constituent contributions from top and bottom
loops are denoted by the 
dashed black and red lines, respectively,
and the black solid lines are for the total sum. 
}
\label{fig:dcu}
\end{figure}
In Fig.~\ref{fig:dcu}, we show the absolute values of the
up-quark CEDM as a function of $C_v$ in units of ${\rm cm}$
with $C_u^S=C_u^P=1/2$ for Types I -- IV of 2HDMs. 
The dashed black and red lines
are for the top- and bottom-loop contributions, 
and the black solid line for the total sum.
Since the Barr--Zee diagrams 
contributing to the up-quark CEDM are dominated by
the top-quark loops, the black dashed lines almost 
overlap with black solid lines.
Note that the top contributions are proportional to 
\begin{equation}
\left(d^C_u\right)_{\rm I,II,III,IV} \propto
-\left[f(\tau_{tH})+g(\tau_{tH})\right]\, C_u^S\,C_u^P\,,
\end{equation}
independent of the 2HDM
types and of $C_v$: see Eq.~(\ref{eq:cedm}).

\begin{figure}[t!]
\hspace{ 0.0cm}
\vspace{-0.5cm}
\centerline{\epsfig{figure=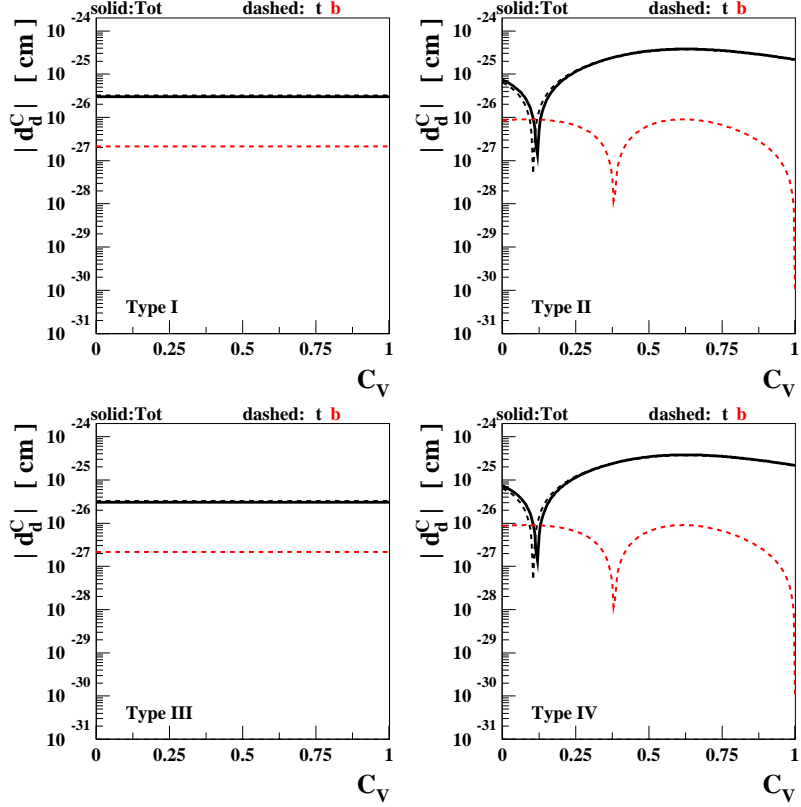,height=12cm,width=12cm}}
\vspace{-0.5cm}
\caption{\it 
The same as in Fig.~\ref{fig:dcu} but for the down-quark CEDM.
}
\label{fig:dcd}
\end{figure}
In Fig.~\ref{fig:dcd}, we show the absolute values of the
down-quark CEDM as a function of $C_v$ in units of ${\rm cm}$
with $C_u^S=C_u^P=1/2$ for Types I -- IV of 2HDMs. The labeling of lines
is the same as in Fig.~\ref{fig:dcu}.
The dominant top-quark
loop contributions are proportional to
\begin{eqnarray} 
\left(d_d^C\right)_{\rm I,III} &\propto & -
\left[-f(\tau_{tH})+g(\tau_{tH})\right]\, C_u^S\,C_u^P\,,
\nonumber \\[2mm]
\left(d_d^C\right)_{\rm II,IV} &\propto & 
-\left[t_\beta^2 C_u^S\,f(\tau_{tH})
+\frac{O_{\phi_1 i}}{c_\beta} \,g(\tau_{tH})\right] \,C_u^P
\end{eqnarray}
for Types I,III and II,IV, respectively:
see Eq.~(\ref{eq:cedm}). Therefore, in Types I and III,
the top contributions are independent of $C_v$, while  
in Types II and IV there is
cancellation around $C_v=0.1$,
similar to the top-quark contributions to $d^E_d$: see
Fig.~\ref{fig:ded}.

\begin{figure}[t!]
\hspace{ 0.0cm}
\vspace{-0.5cm}
\centerline{\epsfig{figure=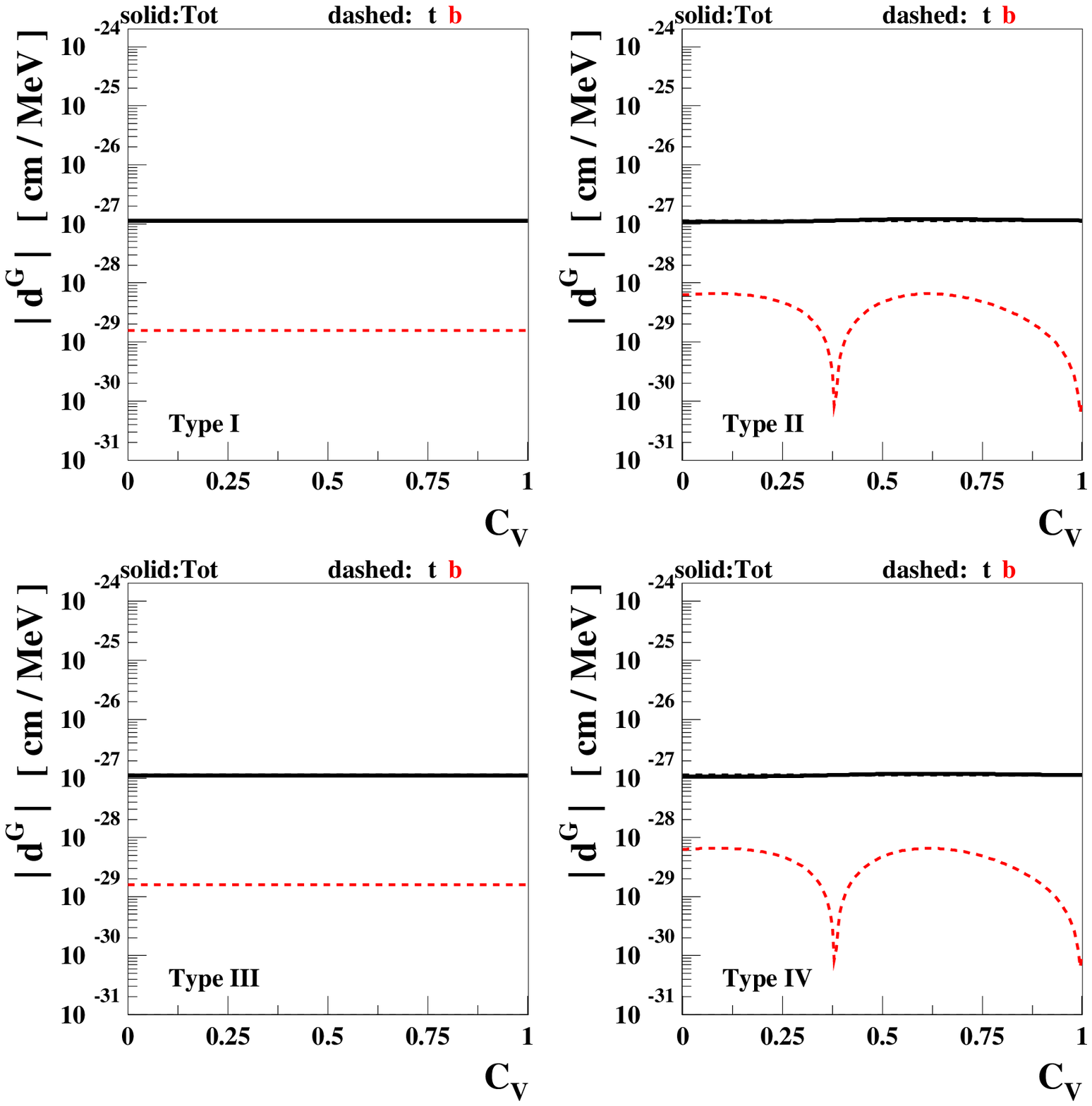,height=12cm,width=12cm}}
\vspace{-0.5cm}
\caption{\it 
The same as in Fig.~\ref{fig:dcu} but for the coefficient
of the Weinberg operator in units of ${\rm cm}/{\rm MeV}$.
}
\label{fig:dgw}
\end{figure}
In Fig.~\ref{fig:dgw}, we show the absolute value of the
coefficient of the Weinberg operator
as a function of $C_v$ in units of ${\rm cm}/{\rm MeV}$
with $C_u^S=C_u^P=1/2$ for Types I -- IV of 2HDMs. The labeling of lines
is the same as in Fig.~\ref{fig:dcu}. 
Again, the dominant contributions are from top loops which are proportional
to
\begin{equation}
\left(d^G\right)_{\rm I,II,III,IV} \propto \, C_u^S\,C_u^P
\end{equation}
and, accordingly, they are independent of $C_v$.

Before closing this subsection, we offer the following comments on the sizes of
(C)EDMs of the light quarks and electron, and $d^G$.
\begin{itemize}
\item{} $|d^E_e|\sim 10^{-27}$ - $10^{-26}\,e\,{\rm cm}$ may induce
$|d_{\rm Tl}|/d_{\rm Tl}^{\rm EXP} \sim 1$,
$|d_{\rm ThO}/{\cal F}_{\rm ThO}|/d_{\rm ThO}^{\rm EXP} \sim {\cal O}(10)$, and
$|d_{\rm Hg}|/d_{\rm Hg}^{\rm EXP} \sim {\cal O}(1)$:
see Eqs.~(\ref{eq:dTl}), (\ref{eq:dThO}), and (\ref{eq:dHgI}).
\item{} $|d^E_u|\sim 10^{-26}\,e\,{\rm cm}$ may induce
$|d_{\rm n}|/d_{\rm n}^{\rm EXP} \sim 10^{-1}$:
see Eq.~(\ref{eq:dnQCD}).
\item{} $|d^E_d|\sim 10^{-26}\,e\,{\rm cm}$ may induce
$|d_{\rm n}|/d_{\rm n}^{\rm EXP} \sim 1$:
see Eq.~(\ref{eq:dnQCD}).
\item{} $|d^C_{u,d}|\sim 10^{-25}\,{\rm cm}$ may induce
$|d_{\rm n}|/d_{\rm n}^{\rm EXP} \sim {\cal O}(1)$ and
$|d_{\rm Hg}^{\rm I}|/d_{\rm Hg}^{\rm EXP} \sim {\cal O}(10)$:
see Eqs.~(\ref{eq:dnQCD}) and (\ref{eq:dHgI}).
\item{} $|d^G|\sim 10^{-27}\,{\rm cm/MeV}$ may induce
$|d_{\rm n}|/d_{\rm n}^{\rm EXP} \sim 6$:
see Eq.~(\ref{eq:dnQCD}).
\end{itemize}
Therefore, the most significant constraints come from 
the thorium-monoxide EDM through $d^E_e$, Mercury EDM through $d^C_{u,d}$,
and neutron EDM through $d^G$.
We are going to present more details in the next subsection.

\subsection{Observable EDMs}
In this subsection, we numerically analyze the 
Thallium, thorium-monoxide, neutron, and Mercury EDMs 
together with their constituent contributions, taking the benchmark point
of $C_u^S=C_u^P=1/2$.

\begin{figure}[t!]
\hspace{ 0.0cm}
\vspace{-0.5cm}
\centerline{\epsfig{figure=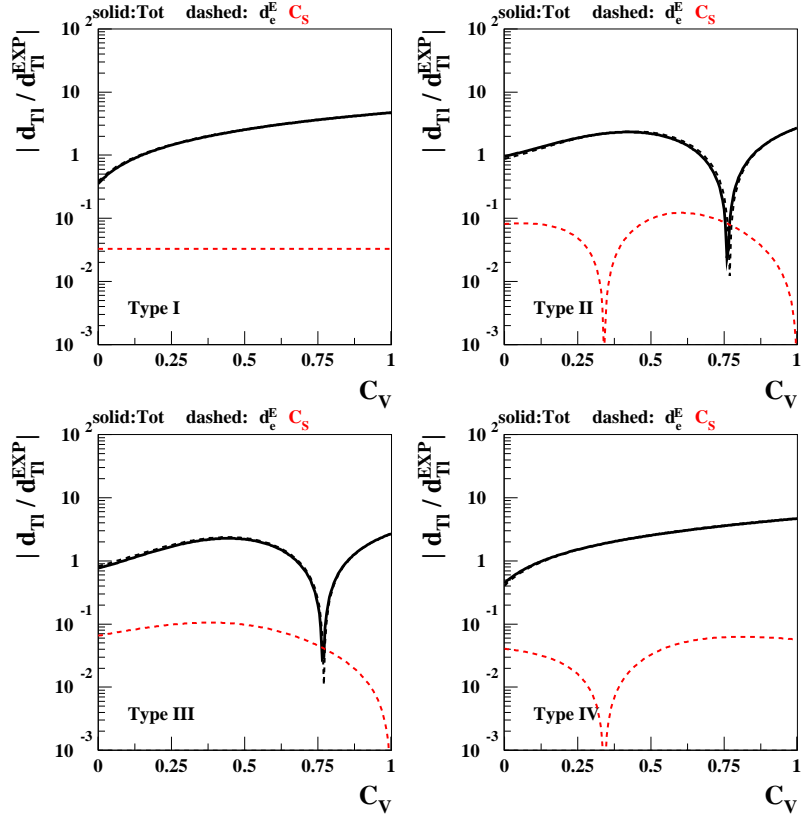,height=12cm,width=12cm}}
\vspace{-0.5cm}
\caption{\it The absolute values of the Thallium EDM as functions of $C_v$
divided by the current experimental limit $d_{\rm Tl}^{\rm EXP}=9\times
10^{-25}\,e\,{\rm cm}$ when $C_u^S=C_u^P=1/2$ for the types I -- IV of 2HDMs. 
The constituent contributions from $d^E_e$ and $C_S$ are denoted by the dashed black
and red lines and the black solid lines are for the total sum.
}
\label{fig:dtl}
\end{figure}
\begin{figure}[t!]
\hspace{ 0.0cm}
\vspace{-0.5cm}
\centerline{\epsfig{figure=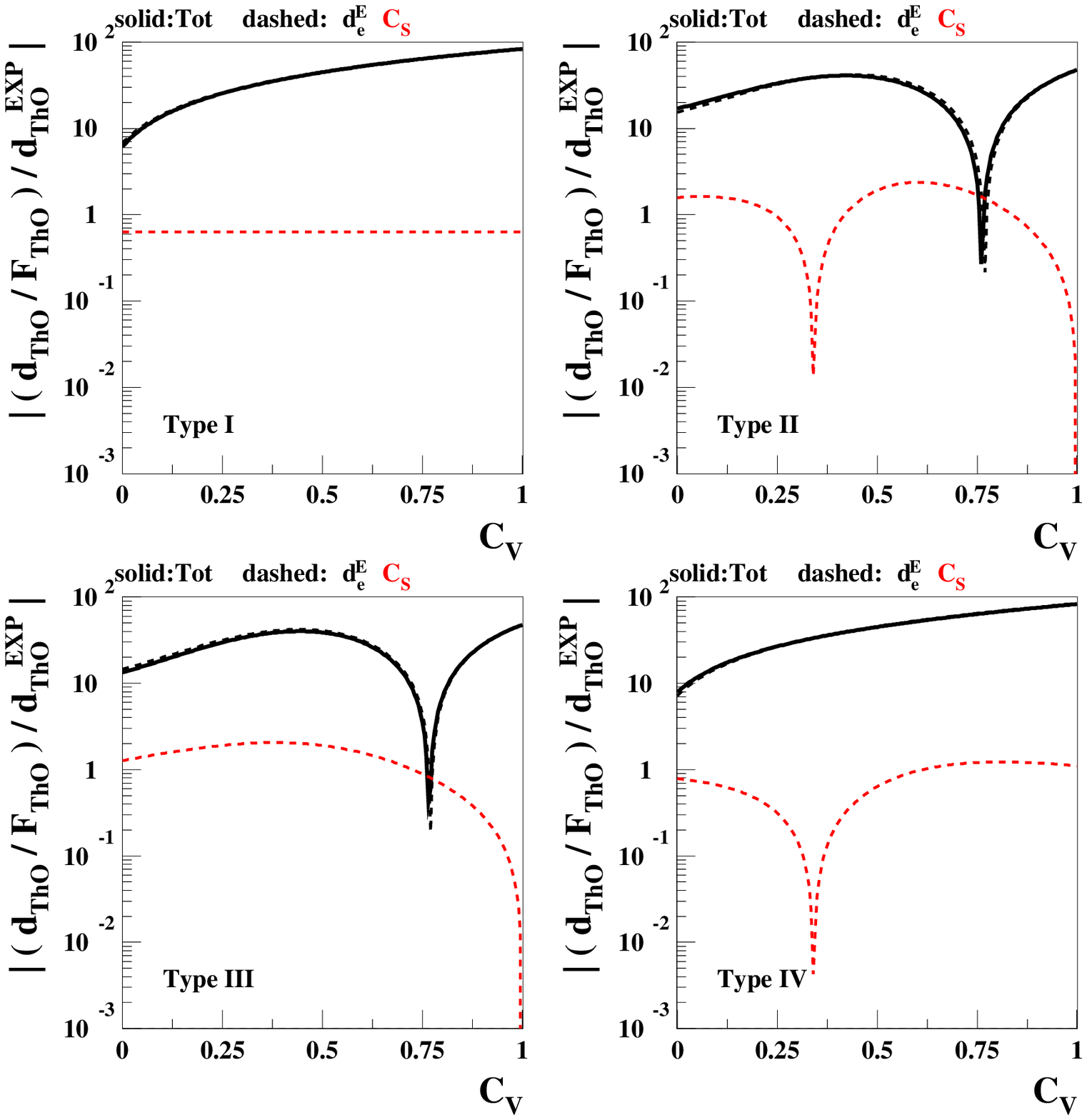,height=12cm,width=12cm}}
\vspace{-0.5cm}
\caption{\it The same as in Fig.~\ref{fig:dtl} but for
the normalized thorium-monoxide EDM $d_{\rm ThO}/{\cal F}_{\rm ThO}$
with $d_{\rm ThO}^{\rm EXP}=8.7\times 10^{-29}\,e\,{\rm cm}$.
}
\label{fig:dtho}
\end{figure}
In Fig.~\ref{fig:dtl},  we show the Thallium EDM 
normalized to the current experimental limit in Eq.~(\ref{eq:exp})
as functions of $C_v$, and in 
Fig.~\ref{fig:dtho} for the normalized thorium-monoxide EDM $d_{\rm ThO}/{\cal
F}_{\rm ThO}$. Both of them are dominated by the electron EDM.
With slightly different subleading $C_S$ contributions,
the behavior  and parametric dependence
of the two EDMs are almost the same:
see Eqs.~(\ref{eq:dTl}) and (\ref{eq:dThO}), 
We observe that the thorium-monoxide EDM indeed
provides one-order of magnitude stronger limits. 
We find $|(d_{\rm ThO}/{\cal F}_{\rm ThO})/d_{\rm ThO}^{\rm EXP}|\lsim 100$ (I, IV)
and $\lsim 50$ (II, III). Moreover, because of the dips near $C_v=0.75$ 
due to the cancellations between the top- and $W$-loop contributions to $d^E_e$
in Types II and III, the thorium-monoxide EDM constraints are shown to
be weaker in Types II and III.
It is interesting to note that
the thorium-monoxide EDM even shows a sensitivity to the $C_S$ contribution.

\begin{figure}[t!]
\hspace{ 0.0cm}
\vspace{-0.5cm}
\centerline{\epsfig{figure=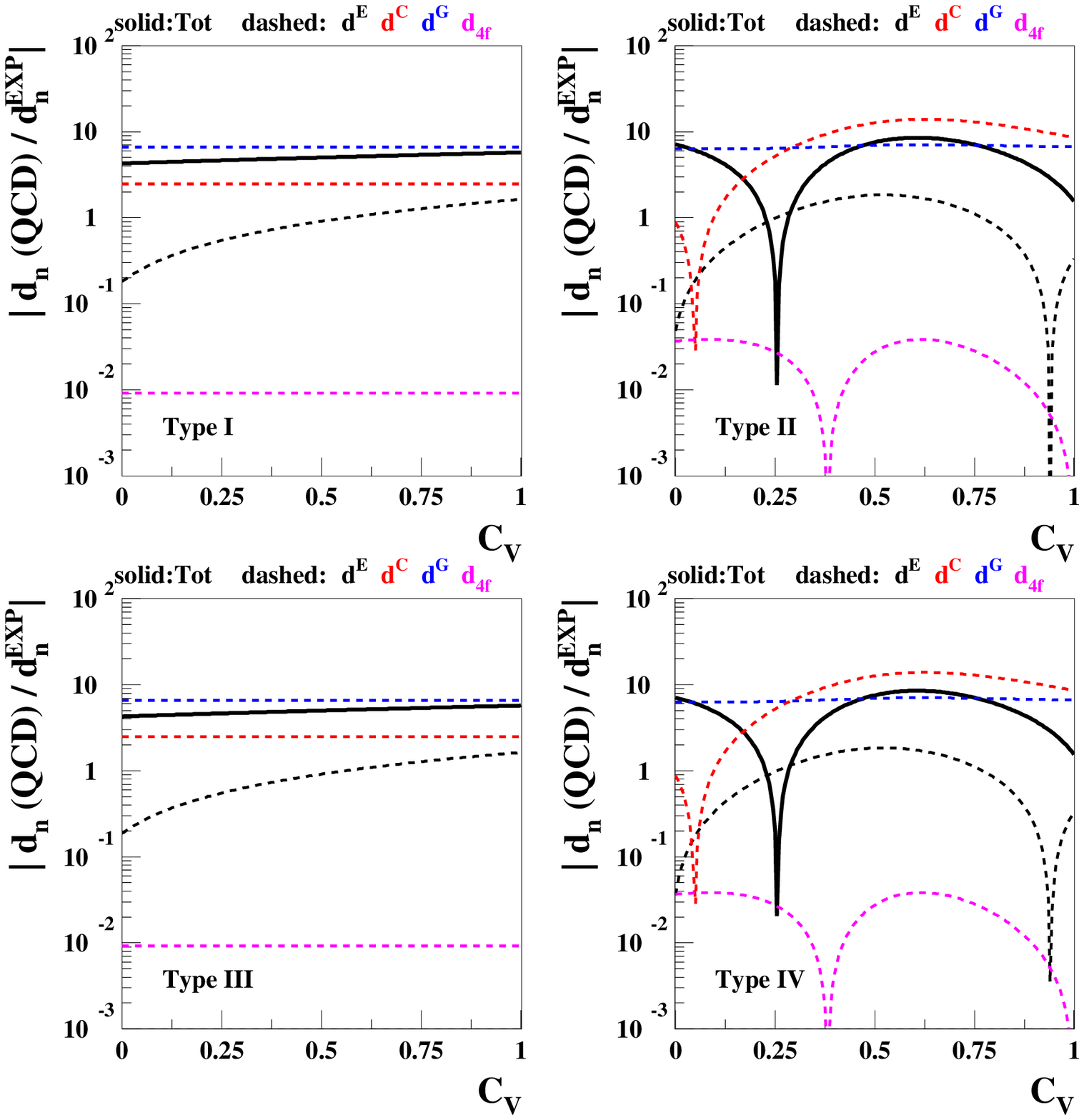,height=12cm,width=12cm}}
\vspace{-0.5cm}
\caption{\it The absolute values of the neutron EDM in
the QCD sum-rule approach as functions of $C_v$
divided by the current experimental limit $d_{\rm n}^{\rm EXP}=2.9\times
10^{-26}\,e\,{\rm cm}$ when $C_u^S=C_u^P=1/2$ for the types I -- IV of 2HDMs. 
The constituent contributions from $d^E_{u,d}$, $d^C_{u,d}$, $d^G$, and 
the four-fermion operators ($d_{4f}$) are denoted by the dashed black, red, blue,
and magenta lines.  The black solid lines are for the total sum.
}
\label{fig:dnQCD}
\end{figure}
Figure~\ref{fig:dnQCD} shows the neutron EDM (black sold lines)
and its constituent contributions from
$d^E_{u,d}$, $d^C_{u,d}$, $d^G$, and
the four-fermion operators as functions of $C_v$  taking $C_u^S=C_u^P=1/2$. 
We observe $|d_{\rm n}/d_{\rm n}^{\rm EXP}|\lsim 10$.
We also
observe the $d^C_{u,d}$ (red dashed lines) and $d^G$ (blue dashed lines)
contributions dominate and they have opposite signs to each other
except for 
the regions near $C_v=0$ in Types II and IV. The cancellation between
the $d^C_{u,d}$ and $d^G$ contributions is most prominent at $C_v=0.25$ 
in Types II and IV, but the milder
cancellation around $C_v=1$ is phenomenologically more important 
because the current Higgs data prefer
the region around $C_v=1$. The cancellation around $C_v=1$ makes the
neutron EDM constraints in Types II and IV weaker than in Types I and III,
as shown in Fig.~\ref{fig:dnQCD}.
We note that, in Types I and III the neutron EDM also show a sensitivity to
the $d^E_{u,d}$ EDMs (black dashed lines) near $C_v=1$.

\begin{figure}[t!]
\hspace{ 0.0cm}
\vspace{-0.5cm}
\centerline{\epsfig{figure=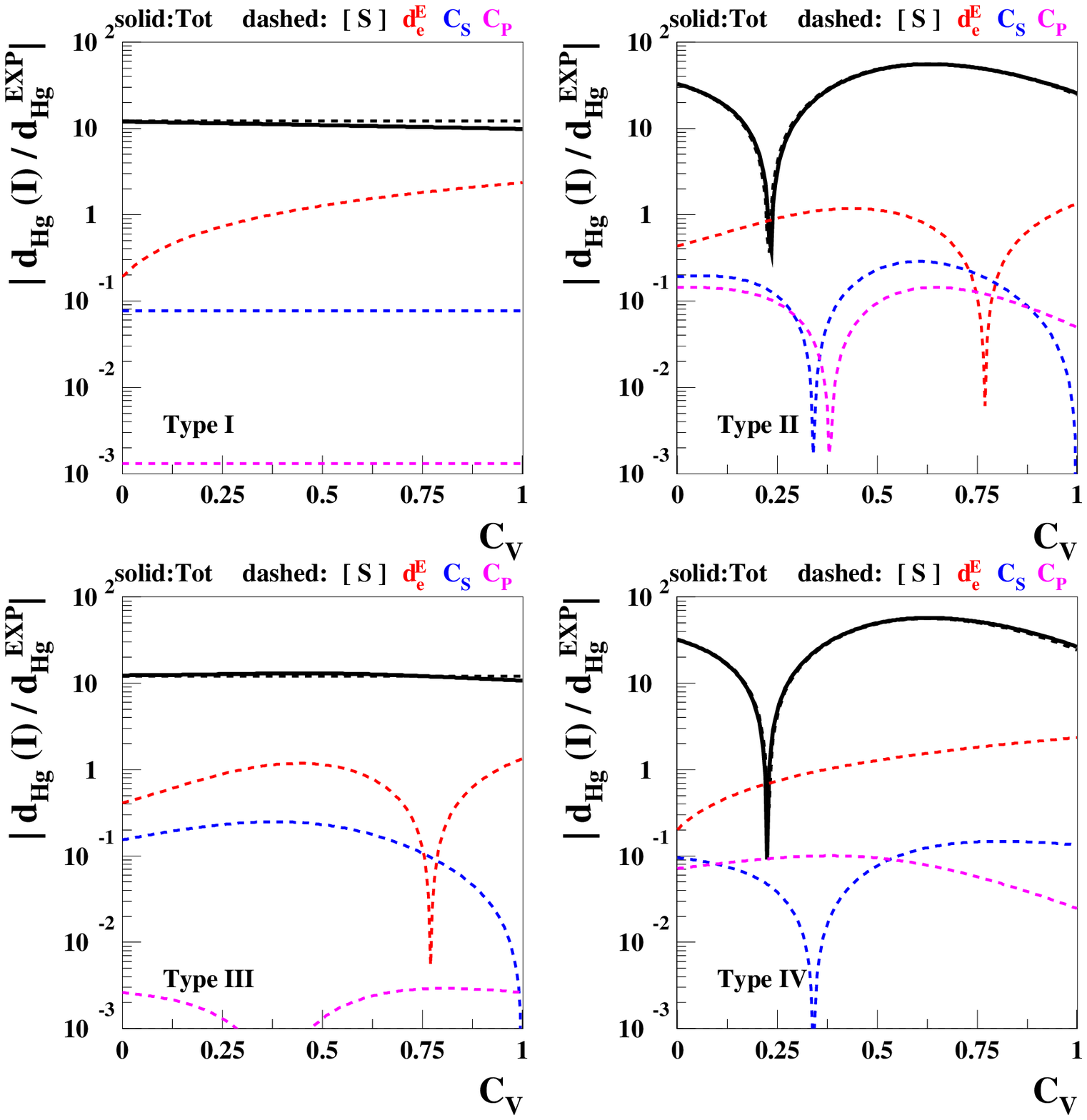,height=12cm,width=12cm}}
\vspace{-0.5cm}
\caption{\it The absolute values of the Mercury EDM using $d_{\rm Hg}^{\rm I}[S]$
as functions of $C_v$
divided by the current experimental limit $d_{\rm Hg}^{\rm EXP}=3.1\times
10^{-29}\,e\,{\rm cm}$ when $C_u^S=C_u^P=1/2$ for Types I -- IV of 2HDMs. 
The constituent contributions from the Schiff moment,
$d^E_e$, $C_S$, and $C_{P}^{(\prime)}$
are denoted by the dashed black, red, blue,
and magenta lines.  The black solid lines are for the total sum.
}
\label{fig:dhg1}
\end{figure}
Figure~\ref{fig:dhg1} shows the Mercury EDM (black sold lines) using $d_{\rm
Hg}^{\rm I}[S]$ for the Schiff moment
and its constituent contributions from
the Schiff moment, $d^E_e$, $C_S$, and $C_{P}^{(\prime)}$
as functions of $C_v$  taking $C_u^S=C_u^P=1/2$. 
We observe $|d_{\rm Hg}/d_{\rm Hg}^{\rm EXP}|\approx 10$ (I, III) and $30$ (II, IV)
around $C_v=1$.
The Mercury EDM is dominated by the contributions from
the Schiff moment (dashed black lines) and
has also a sensitivity to
the electron EDM (red dashed lines) near $C_v=1$.

\subsection{EDM Constraints}
In this subsection, we present the CL regions in the $C_u^S$-$C_u^P$ plane
which satisfy 
the current Higgs-boson data and various EDM constraints.

\begin{figure}[th]
\hspace{ 0.0cm}
\vspace{-0.5cm}
\includegraphics[angle=-90,width=7cm]{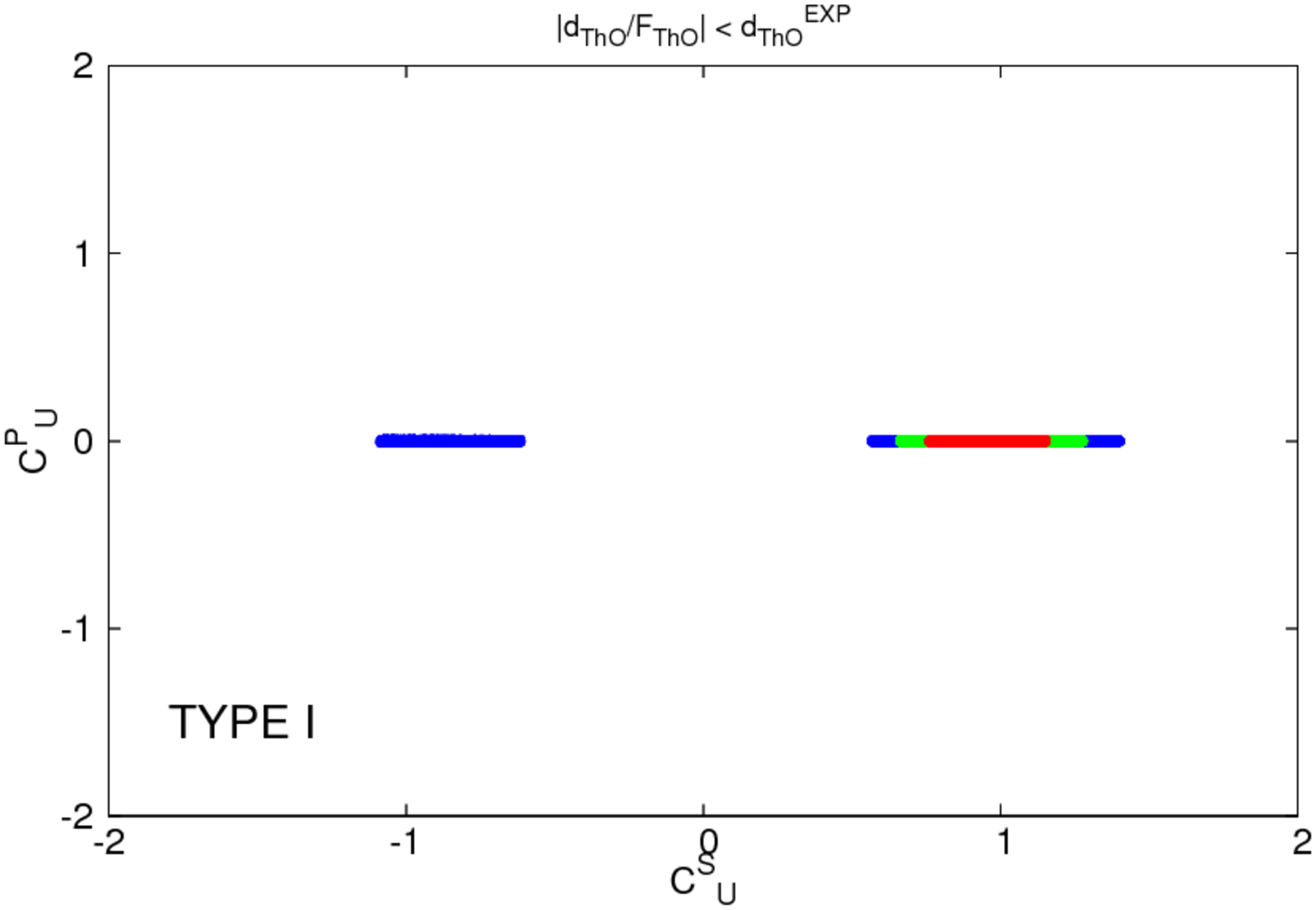}
\includegraphics[angle=-90,width=7cm]{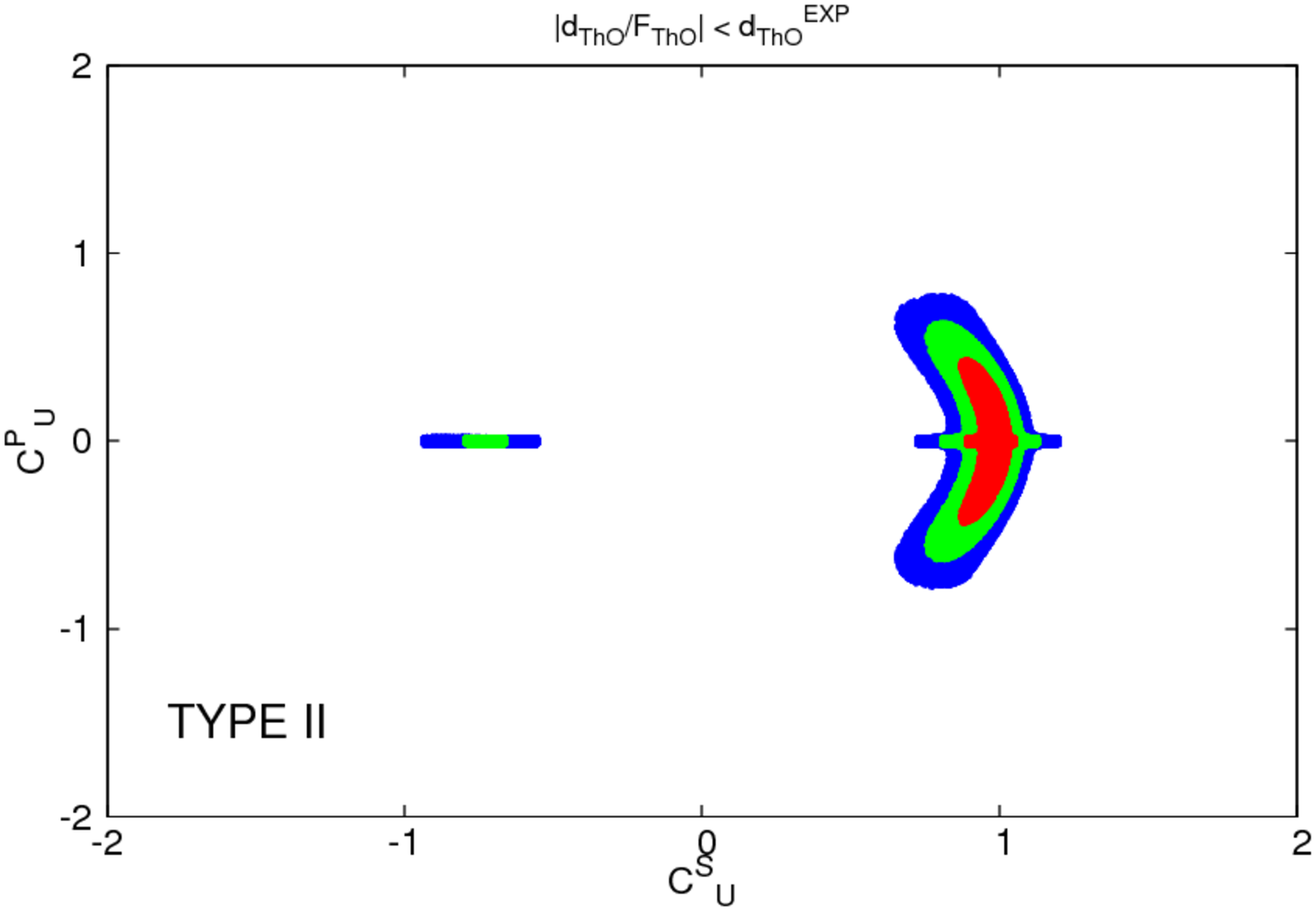}
\\[10mm]
\includegraphics[angle=-90,width=7cm]{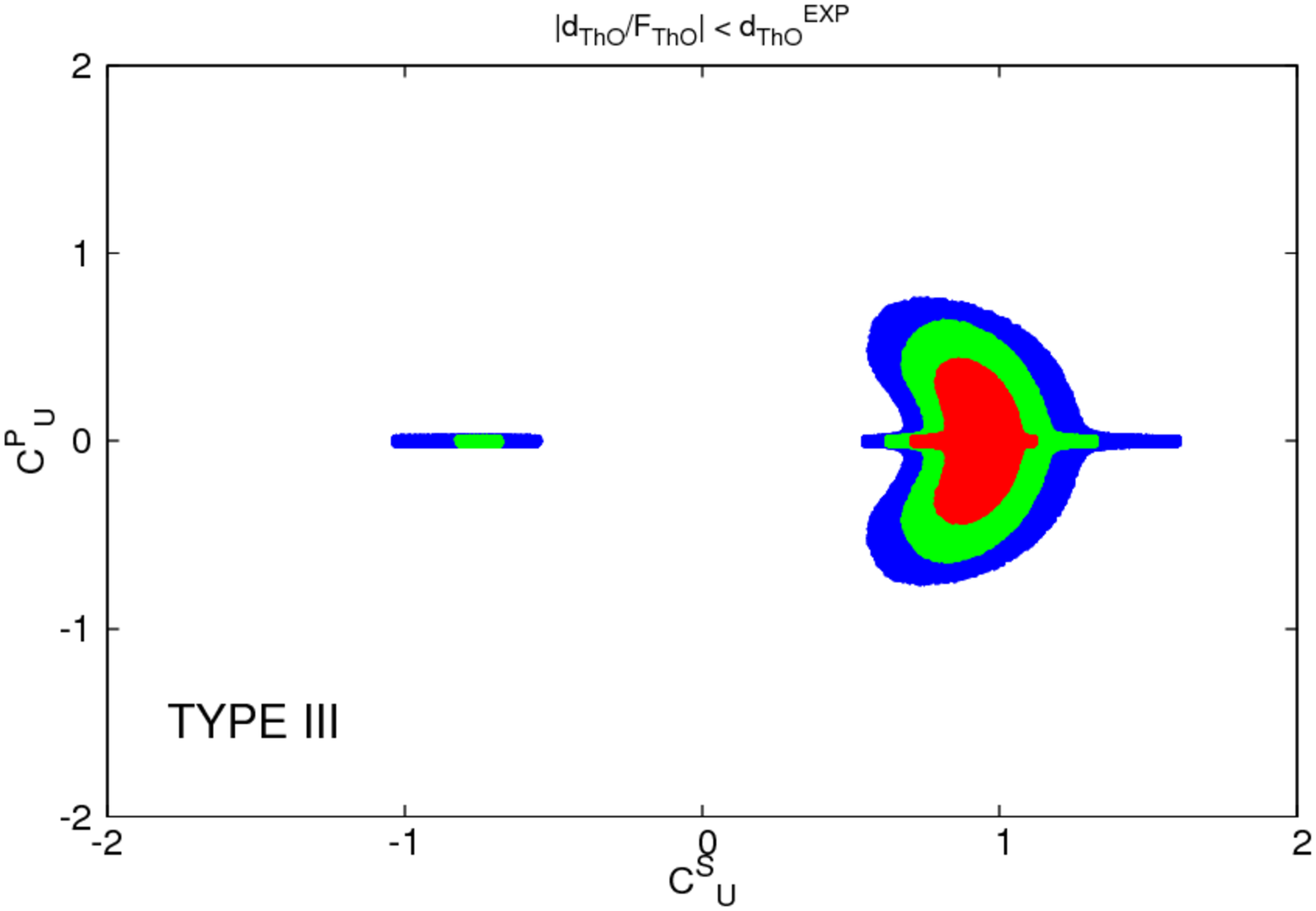}
\includegraphics[angle=-90,width=7cm]{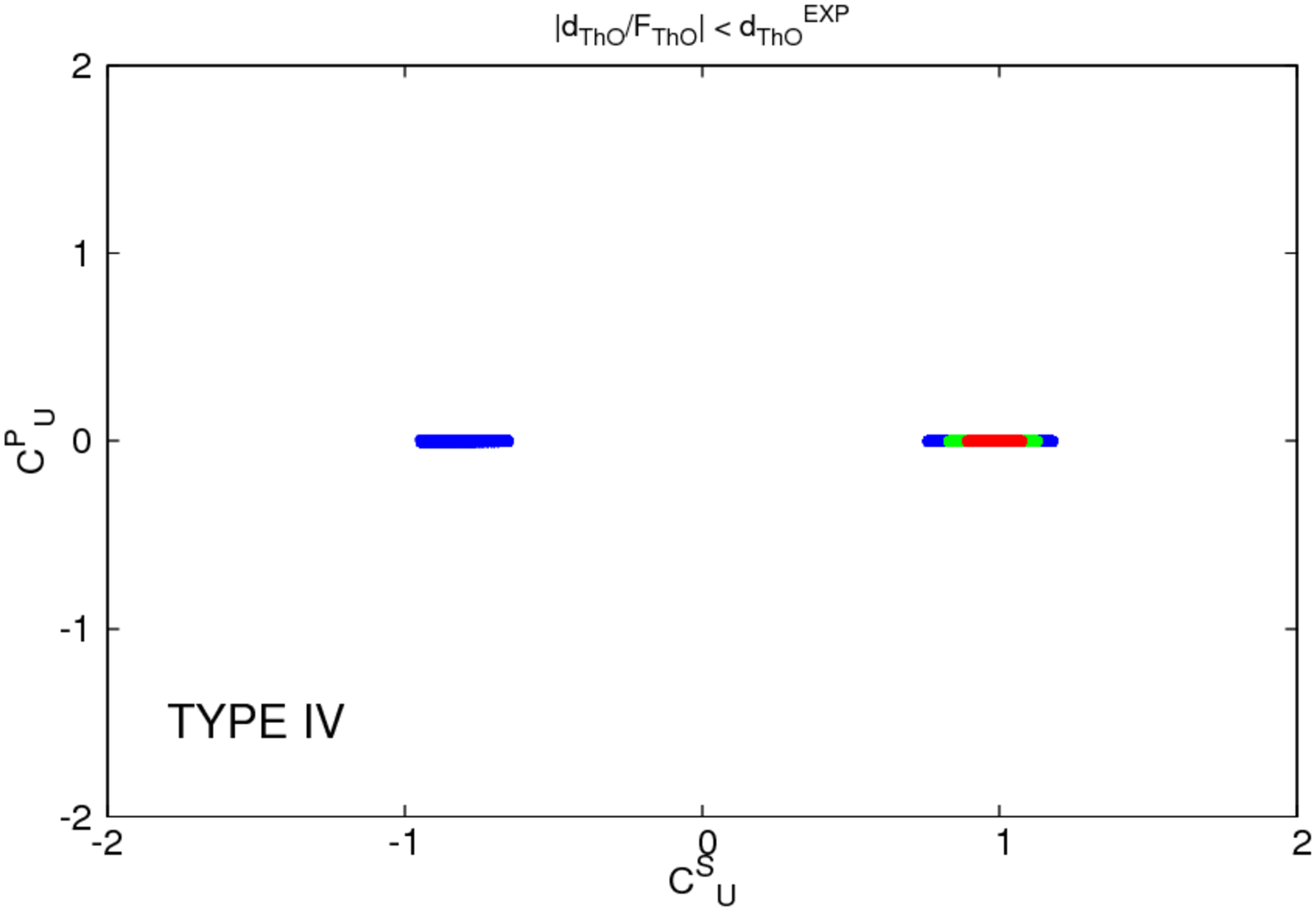}
\vspace{0.5cm}
\caption{\it The same as in Fig.~\ref{fig:hfit} but
with the thorium-monoxide EDM constraint 
$|(d_{\rm ThO}/{\cal F}_{\rm ThO})/d_{\rm ThO}^{\rm EXP}|\leq 1$ applied.
}
\label{fig:hfit_tho}
\end{figure}
\begin{figure}[th]
\hspace{ 0.0cm}
\vspace{-0.5cm}
\includegraphics[angle=-90,width=7cm]{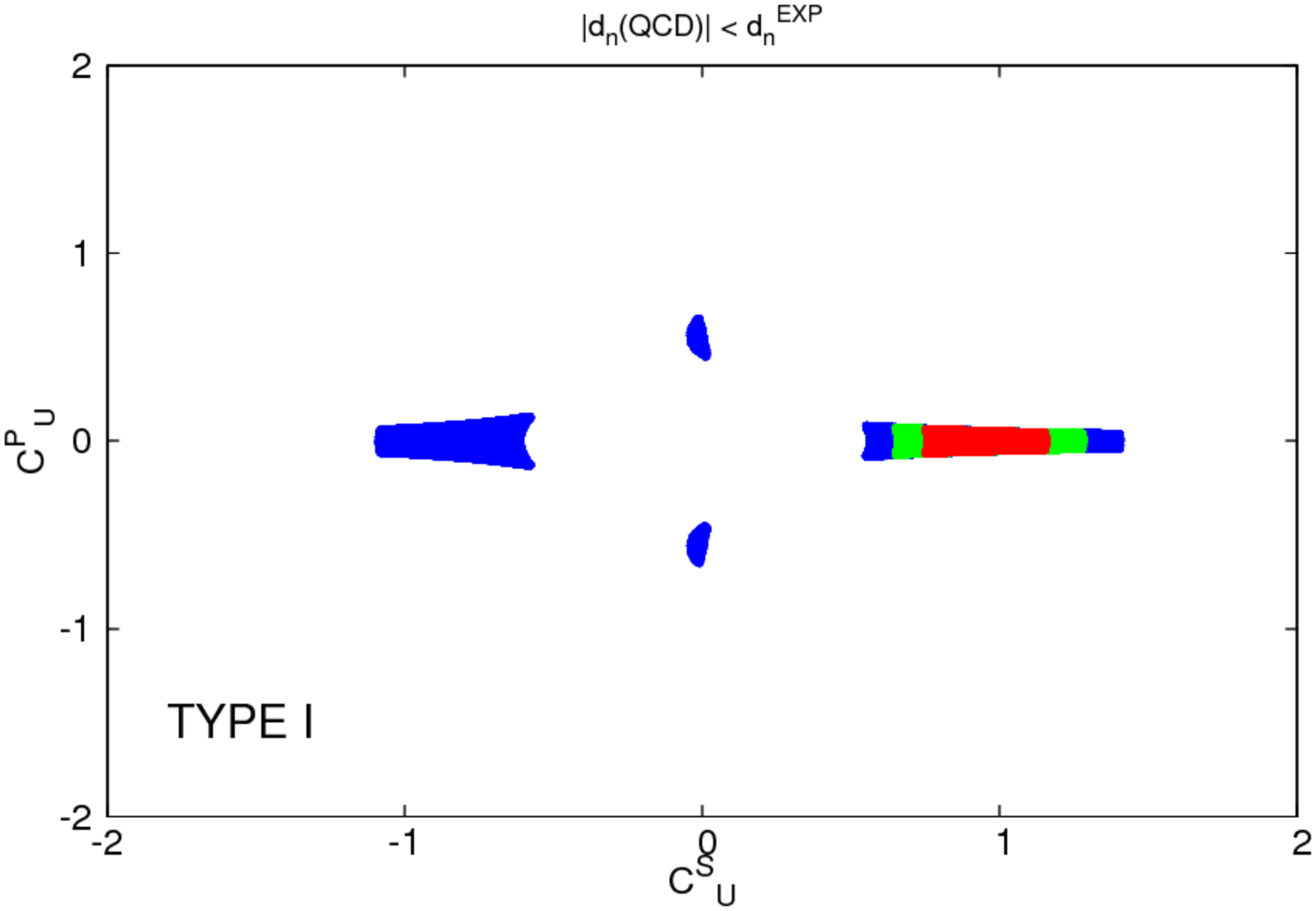}
\includegraphics[angle=-90,width=7cm]{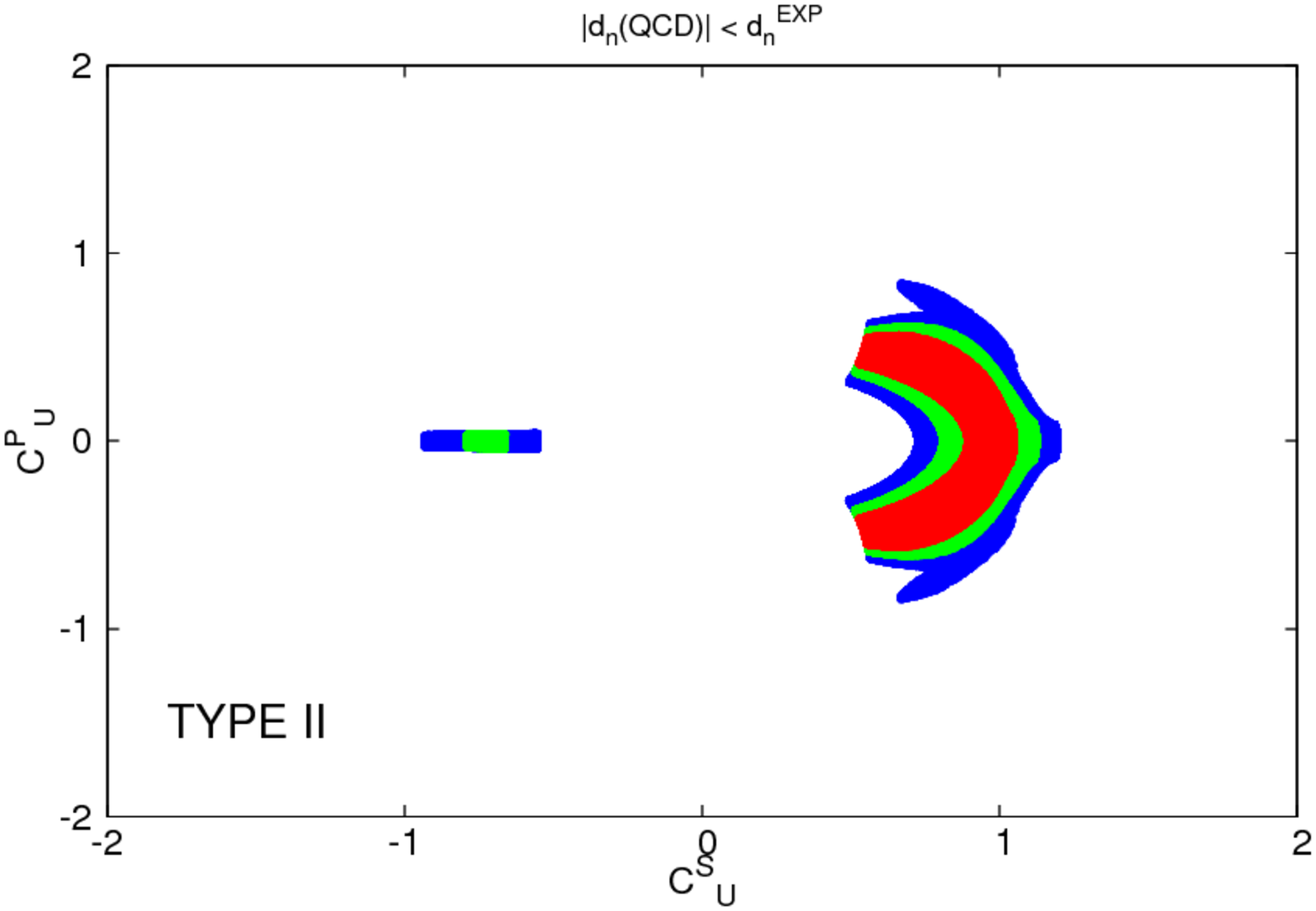}
\\[10mm]
\includegraphics[angle=-90,width=7cm]{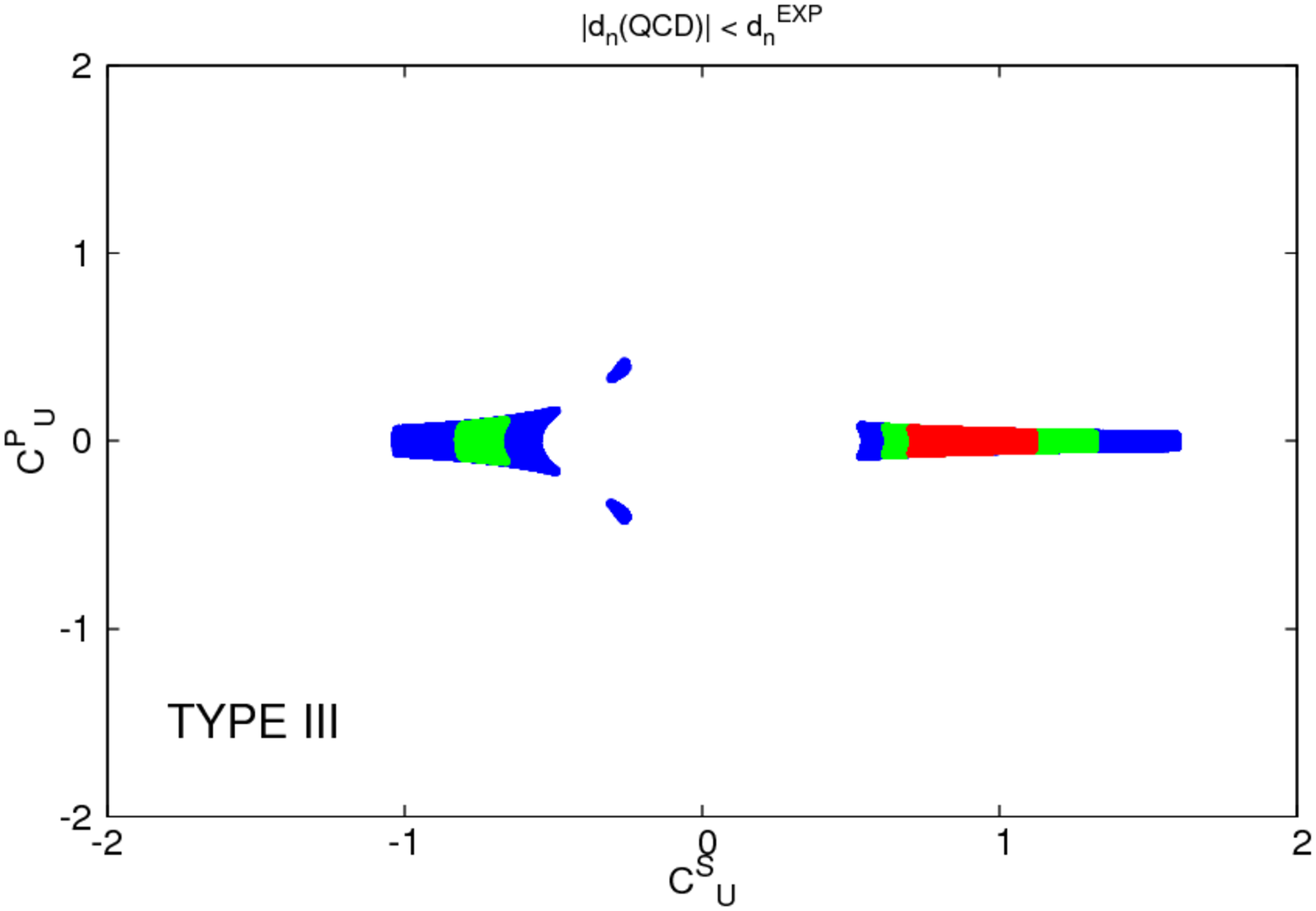}
\includegraphics[angle=-90,width=7cm]{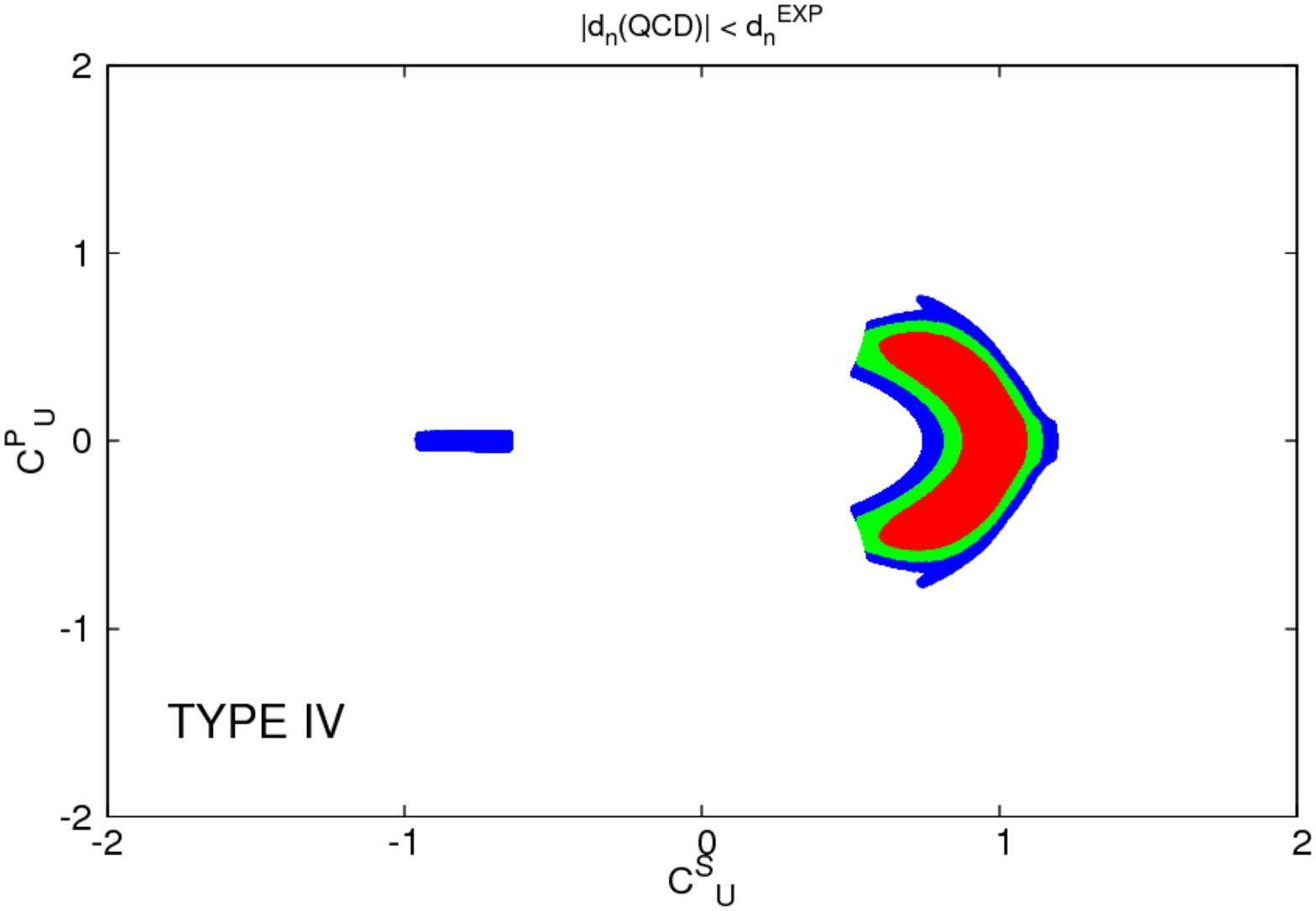}
\vspace{0.5cm}
\caption{\it The same as in Fig.~\ref{fig:hfit} but
with the neutron EDM constraint constraint
$|d_{\rm n}/d_{\rm n}^{\rm EXP}|\leq 1$ applied.
}
\label{fig:hfit_n}
\end{figure}
\begin{figure}[th]
\hspace{ 0.0cm}
\vspace{-0.5cm}
\includegraphics[angle=-90,width=7cm]{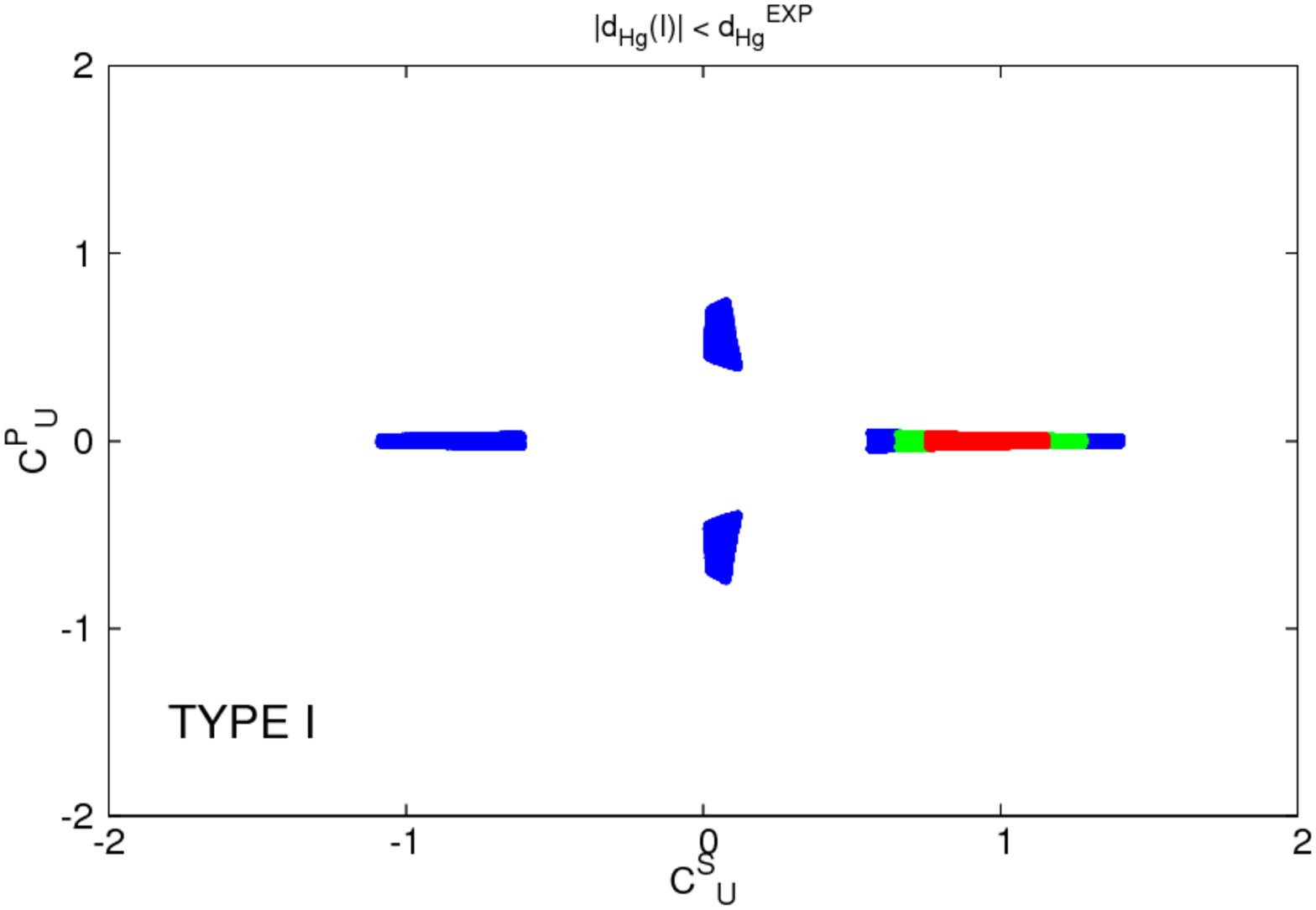}
\includegraphics[angle=-90,width=7cm]{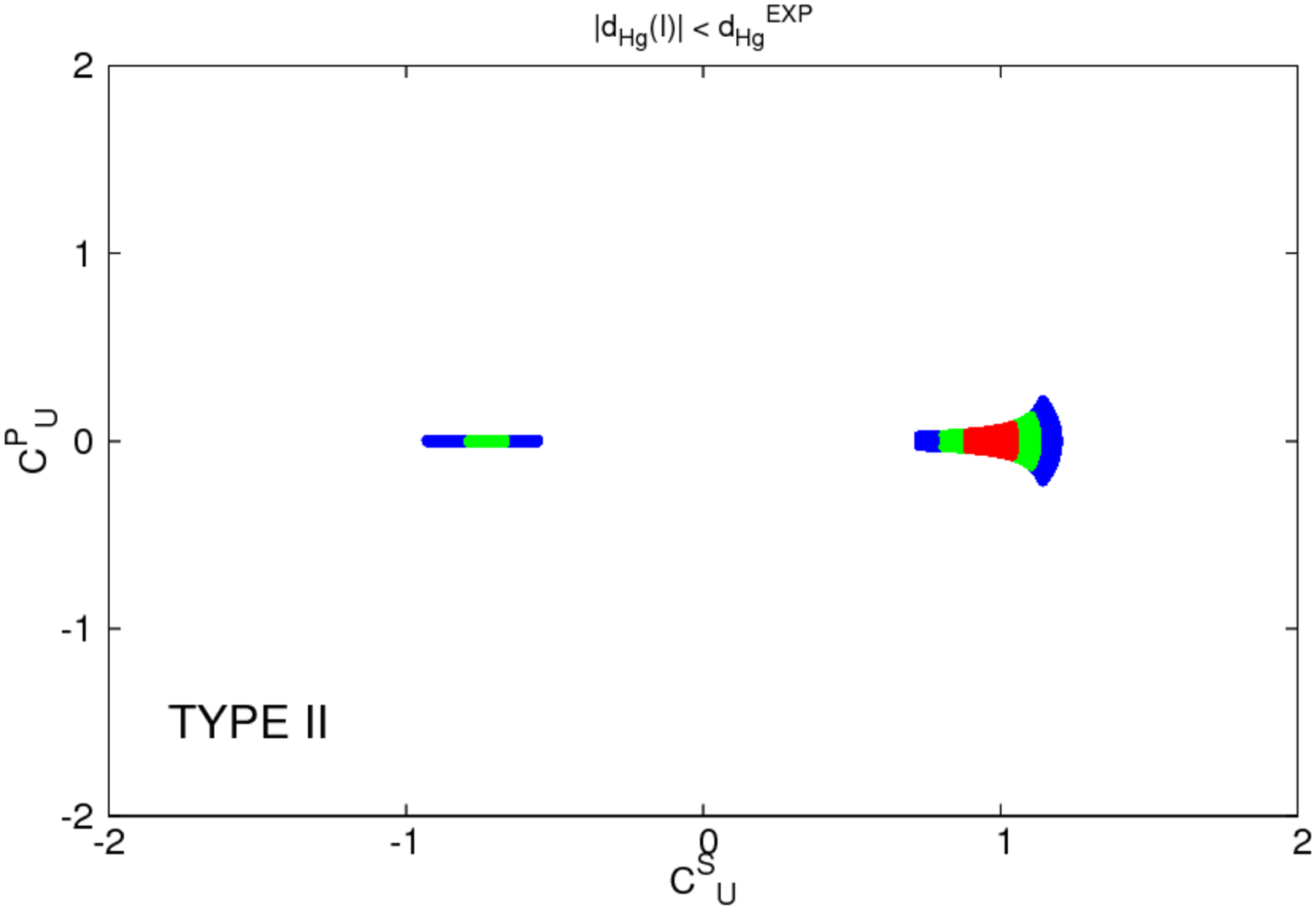}
\\[10mm]
\includegraphics[angle=-90,width=7cm]{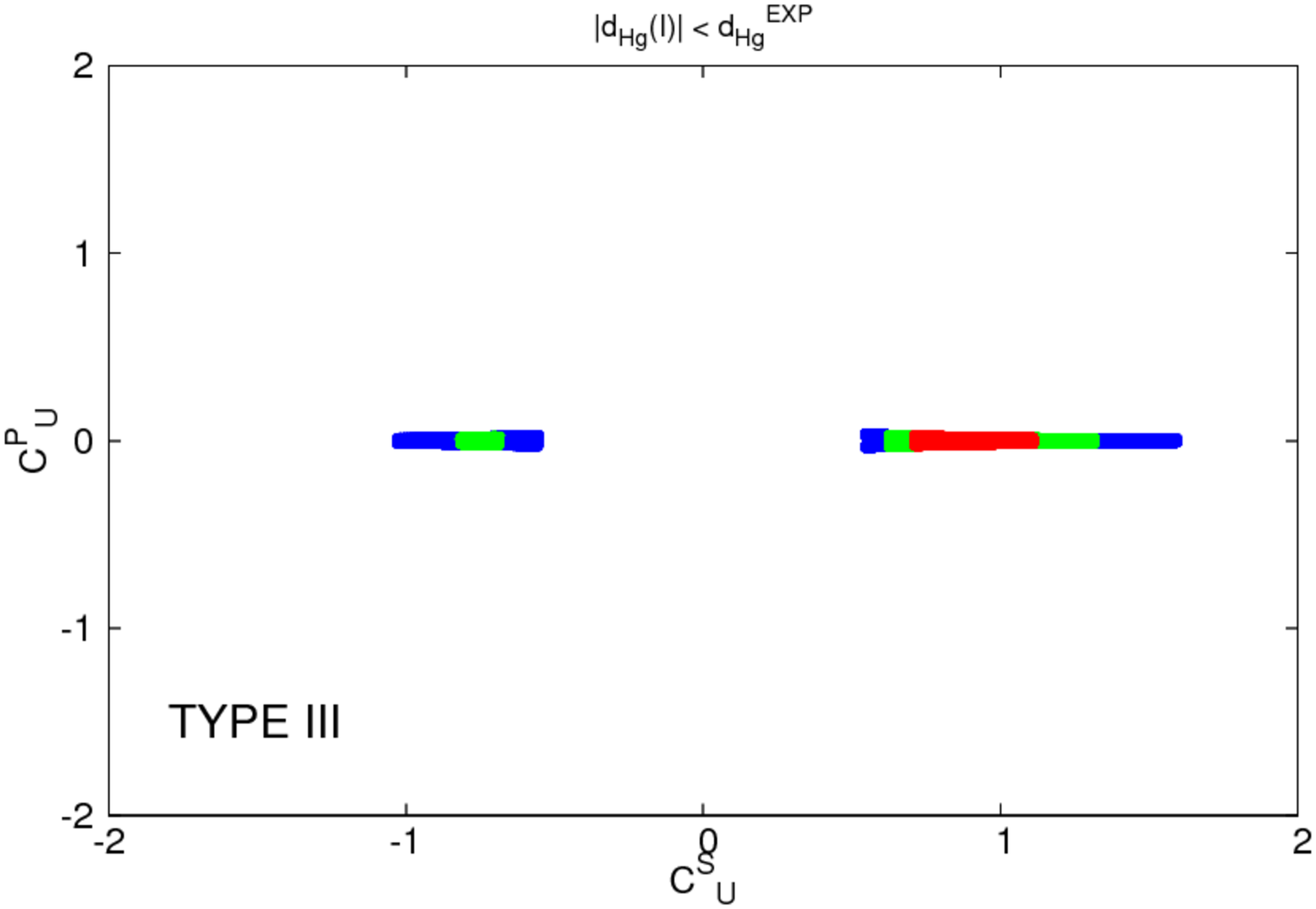}
\includegraphics[angle=-90,width=7cm]{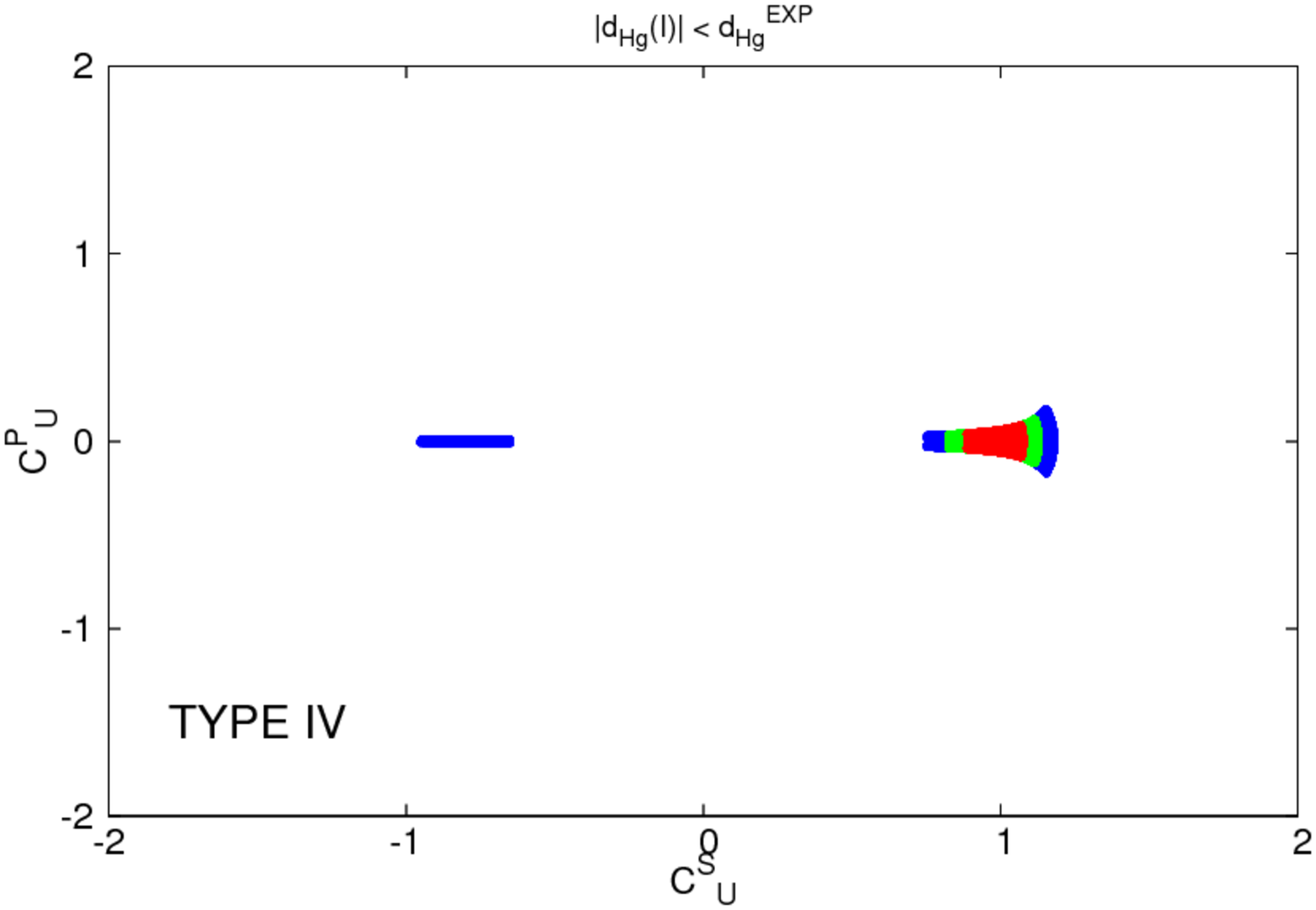}
\vspace{0.5cm}
\caption{\it The same as in Fig.~\ref{fig:hfit} but
with the Mercury EDM constraint
$|d_{\rm Hg}^{\rm I}/d_{\rm Hg}^{\rm EXP}|\leq 1$ applied.
}
\label{fig:hfit_hg}
\end{figure}

In Fig.~\ref{fig:hfit_tho},  we show the allowed regions
satisfying the Higgs-boson data and
the thorium-monoxide EDM constraint
at $68.3\%$ (red), $95\%$ (green), and $99.7\%$ (blue) CL 
in the plane of $C_u^S$ vs $C_u^P$ for Types I -- IV.
We recall that the CL regions before applying the EDM constraints
have been shown in Fig.~\ref{fig:hfit}.
For each allowed 
point in the $C_u^S$-$C_u^P$ plane in Fig.~\ref{fig:hfit},
the thorium-monoxide
EDM is calculated, and we accept the point if
$|(d_{\rm ThO}/{\cal F}_{\rm ThO})/d_{\rm ThO}^{\rm EXP}|\leq 1$ is satisfied while
varying $C_v$ within the corresponding CL regions 
\footnote{We are not showing the
Thallium EDM constraints since they are always weaker than those from the
thorium-monoxide EDM.}.
We observe that $C_u^P\neq 0$ is strongly constrained in Types I and IV.
While in Types II and III,
the constraints are weaker in the regions centered  around the point 
$C_u^S=1$ due to the
cancellation between the top- and $W$-loop contributions to the 
dominant electron
EDM: see Figs.~\ref{fig:dee} and \ref{fig:dtho}. 
We find that $|C_u^P|$ can be as large as
$\sim 0.6$ for Types II and III at $95\%$ CL (green regions).

Figure~\ref{fig:hfit_n} shows the allowed regions
satisfying the Higgs-boson data and the neutron EDM constraint
at $68.3\%$ (red), $95\%$ (green), and $99.7\%$ (blue) CL, respectively,
in the plane of $C_u^S$ vs $C_u^P$ for Type I -- IV. 
The allowed regions are
obtained in the same way as in the case of thorium-monoxide. The neutron EDM
constraint is weaker in Types II and IV due to the cancellation between
the $d^C_{u,d}$ and $d^G$ contributions around $C_v=1$: see Fig.~\ref{fig:dnQCD}.
We find that $|C_u^P|$ can be as large as
$\sim 0.6$ for Types II and IV at $95\%$ CL (green regions).

Figure~\ref{fig:hfit_hg} is the same as in Figures~\ref{fig:hfit_tho} and
\ref{fig:hfit_n} but with the Mercury EDM constraint applied.
In contrast to the weaker thorium-monoxide (neutron) EDM constraint
in Types II and III (Types II and IV),
the Mercury EDM constraint
is almost equally stringent in all four types and, specifically, 
$|C_u^P|$ is restricted
to be $\sim 0.1$ for Types II and IV.

\begin{figure}[th]
\hspace{ 0.0cm}
\vspace{-0.5cm}
\includegraphics[angle=-90,width=7cm]{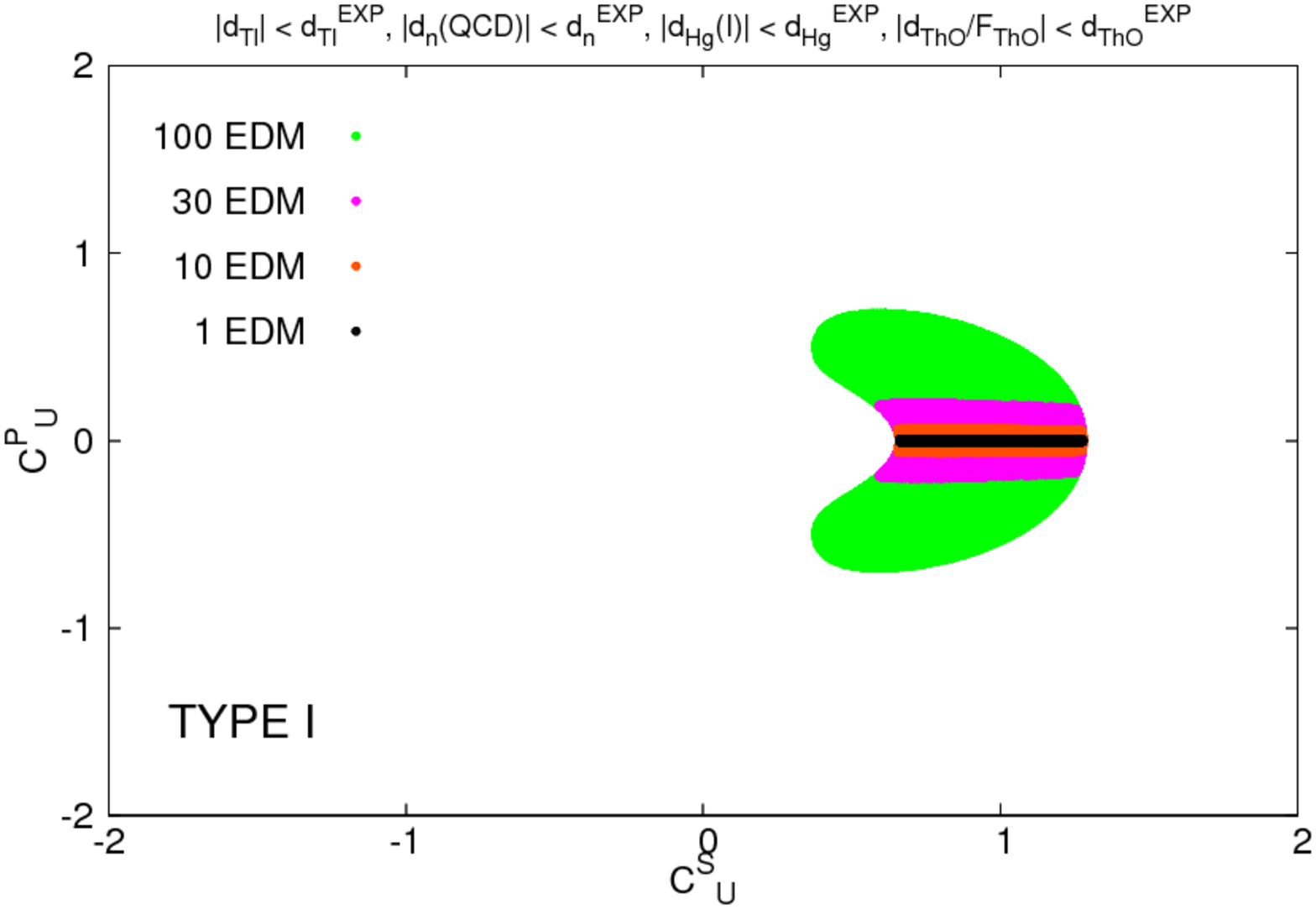}
\includegraphics[angle=-90,width=7cm]{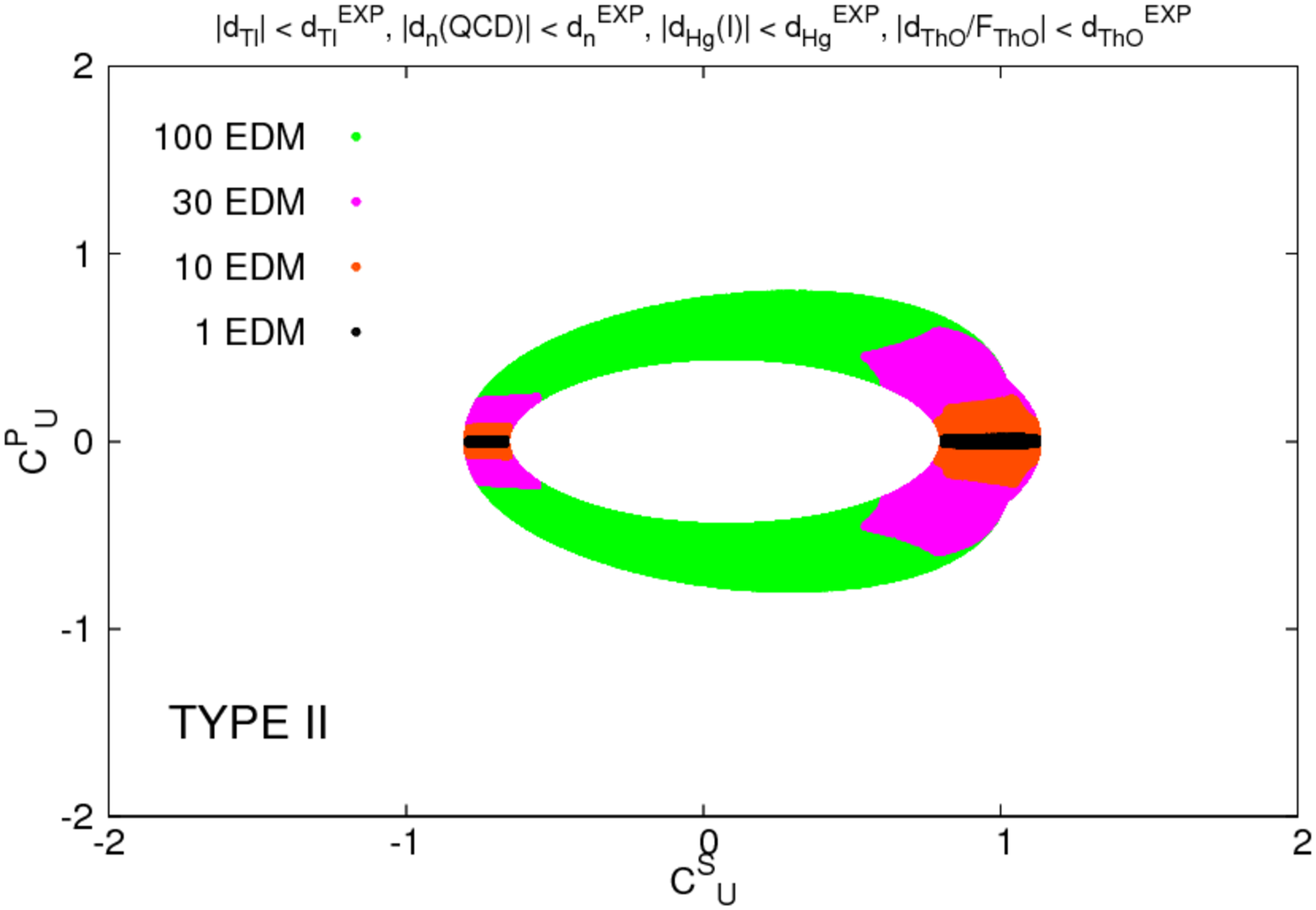}
\\[10mm]
\includegraphics[angle=-90,width=7cm]{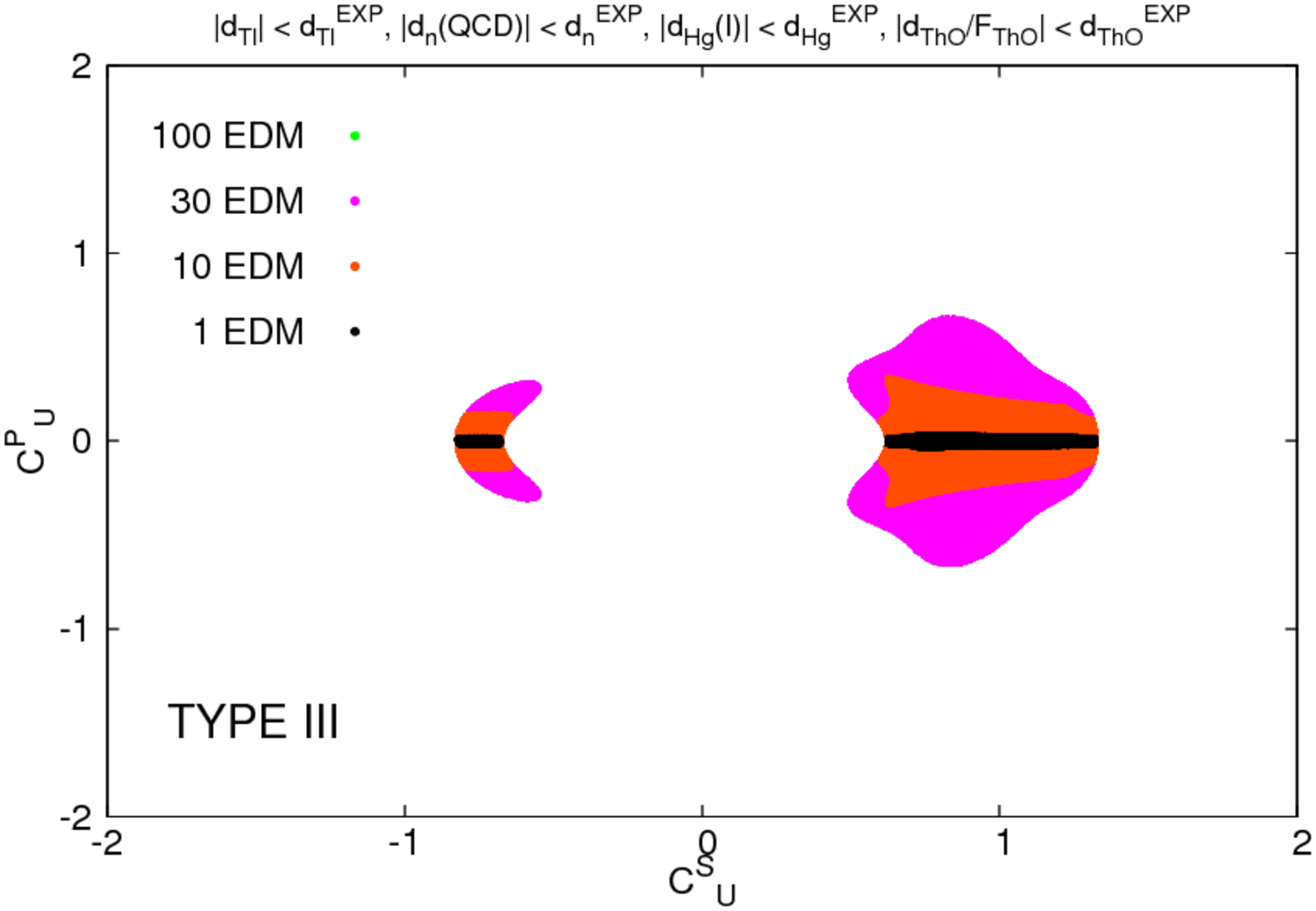}
\includegraphics[angle=-90,width=7cm]{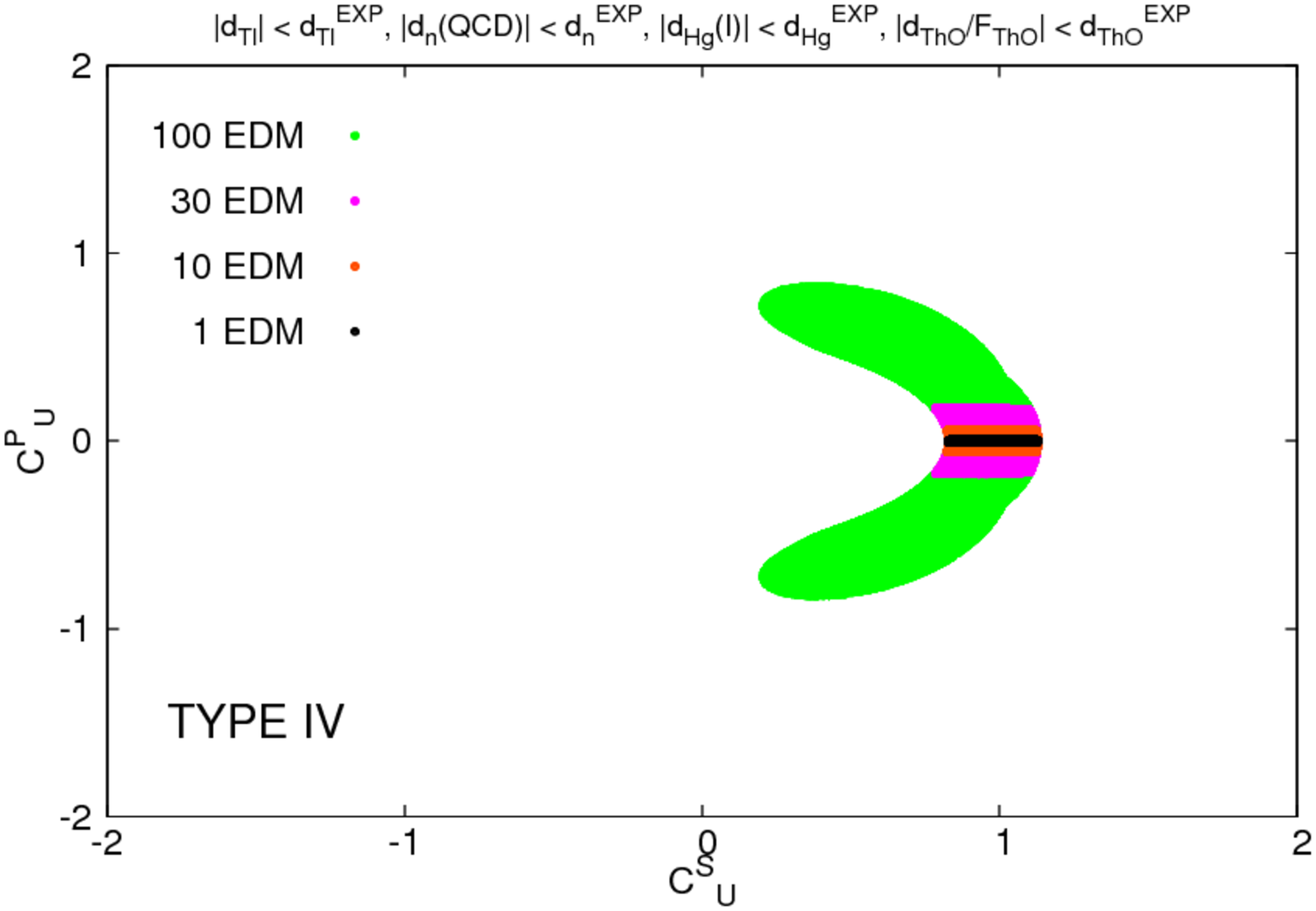}
\vspace{0.5cm}
\caption{\it The $95\,\%$ CL regions satisfying
the Thallium, thorium-monoxide, neutron, and Mercury EDM constraints (black)
simultaneously, as well as the Higgs data.
The orange, pink, and green regions are for the cases of 
applying relaxed constraints
$|d_{\rm Tl, n, Hg}/d^{\rm EXP}_{\rm Tl, n, Hg}|\leq r$ and
$|(d_{\rm ThO}/{\cal F}_{\rm ThO})/d_{\rm ThO}^{\rm EXP}|\leq r$
with the relaxation factor $r=10$ (orange), $30$ (pink), and $100$ (green).
}
\label{fig:hfit_95}
\end{figure}
The combined constraint at 95\% CL from all the EDMs measurements and the 
Higgs-boson data is obtained in Fig.~\ref{fig:hfit_95}.
The black regions in
Fig.~\ref{fig:hfit_95} shows the $95\,\%$ CL regions satisfying
the Thallium, thorium-monoxide, neutron, and Mercury EDM constraints
simultaneously, as well as the Higgs-boson data.
We find that the combination of all available EDM experiments 
provide remarkably tight bounds on CP violation.  Thus,
non-zero values of $C_u^P$ are stringently restricted as
\begin{equation}
\label{eq:cup-limit}
|C_u^P| \ \lsim \ 7\times 10^{-3}\ {\rm (I)}\,, \ \
2\times 10^{-2}\ {\rm (II)}\,, \ \
3\times 10^{-2}\ {\rm (III)}\,, \ \
6\times 10^{-3}\ {\rm (IV)}\,.
\end{equation}

Since we have only taken into account the $125.5$ GeV Higgs-mediated EDMs, 
there could possibly be other  contributions to the EDMs 
if the $125.5$ GeV Higgs $H$ is
embedded in the models beyond the SM. The additional contributions are model
dependent and, for example, they are induced by the other Higgs bosons 
in the 2HDM framework, from some supersymmetric particles in SUSY models,
etc. One may expect that cancellations may occur between 
the $H$-mediated and these additional contributions.
In this case, the EDM constraints can be relaxed. 
In Fig.~\ref{fig:hfit_95}, we also show the $95\,\%$ CL regions satisfying
the relaxed constraints
\begin{equation}
|d_{\rm Tl, n, Hg}/d^{\rm EXP}_{\rm Tl, n, Hg}|\leq r\, \ \ \ {\rm and} \ \ \
|(d_{\rm ThO}/{\cal F}_{\rm ThO})/d_{\rm ThO}^{\rm EXP}|\leq r
\end{equation}
with the relaxation factor $r=10$ (orange), $30$ (pink), and $100$ (green).
The factor $r$, say $r=100$, represents a fine-tuning of order $10^{-2}$.
If the degree of cancellation is $90\,\%$ ($99\,\%$), 
with $100\,\%$  corresponding to a
complete cancellation, the orange (green) regions with 
$r=10\,(100)$ are allowed.
For $r=10$, $|C_u^P|$ can be as large as $\sim 0.1$ (I), $\sim 0.2$ (II), 
$\sim 0.4$ (III), and $\sim 0.1$ (IV).
When $r=100$, we observe the whole $95\,\%$ CL regions are allowed in 
Types I, II, and IV. 
In Type III, the whole $95\,\%$ CL region is allowed for the smaller $r=30$.

\begin{figure}[th]
\hspace{ 0.0cm}
\vspace{-0.5cm}
\includegraphics[angle=-90,width=7cm]{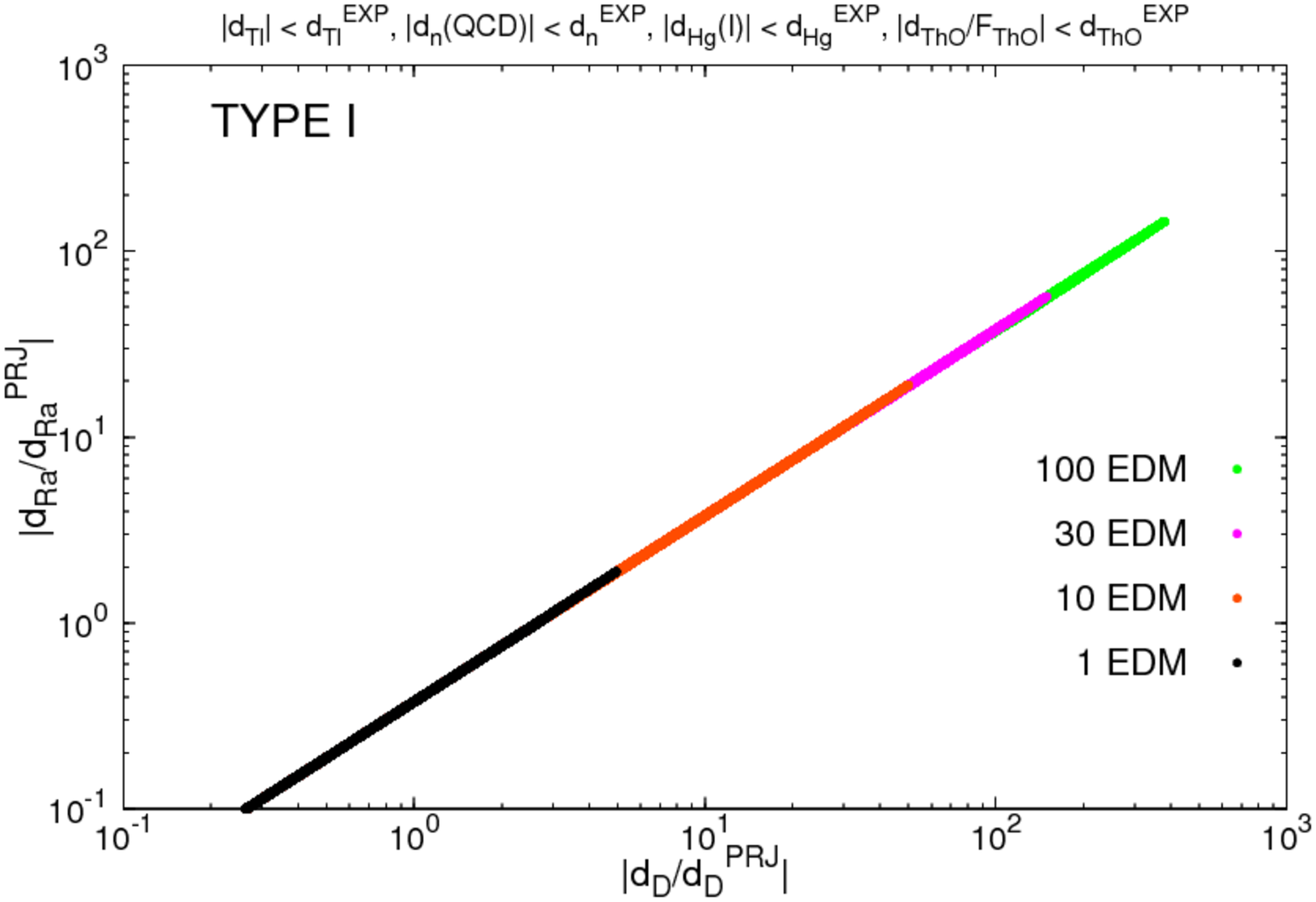}
\includegraphics[angle=-90,width=7cm]{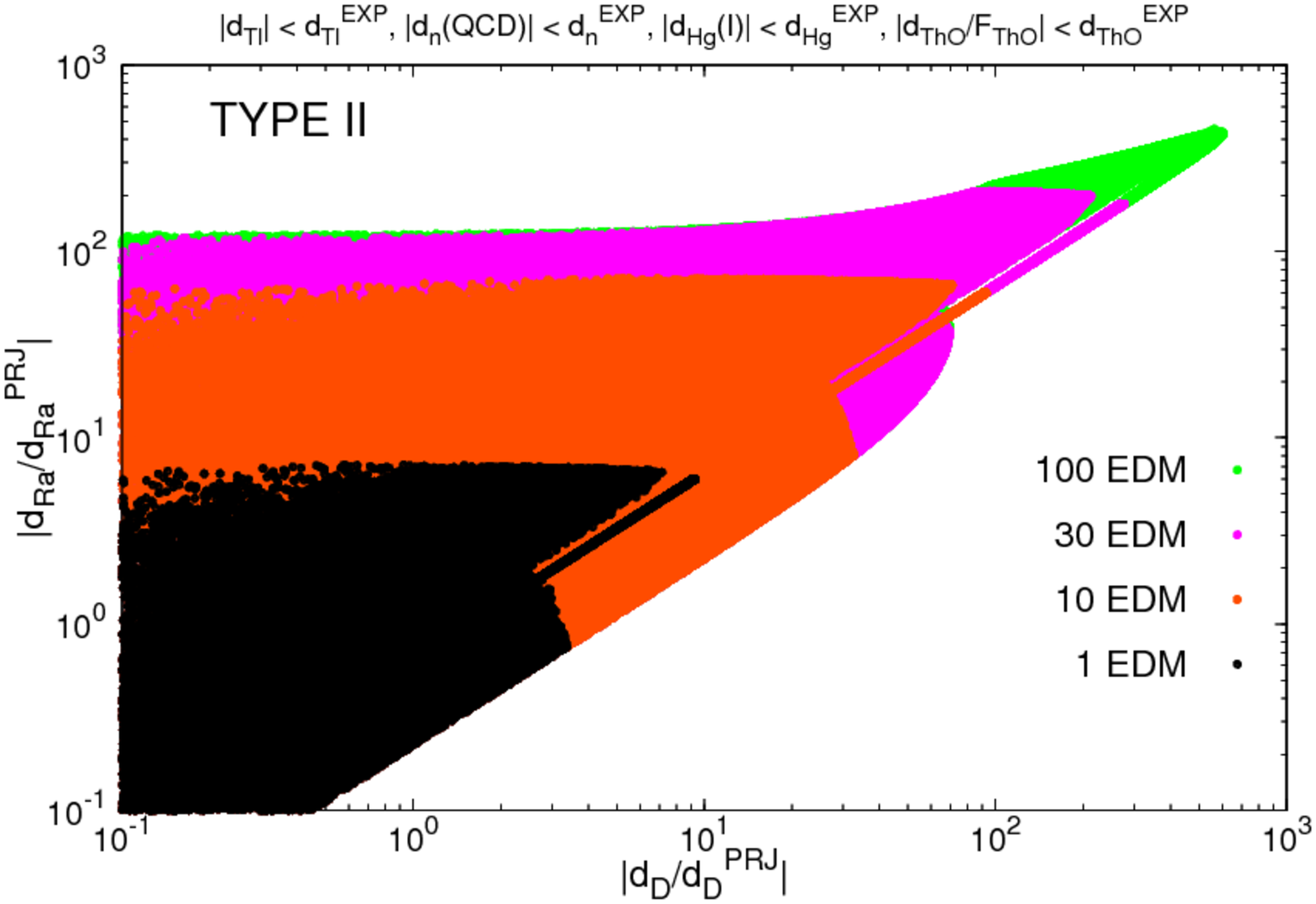}
\\[10mm]
\includegraphics[angle=-90,width=7cm]{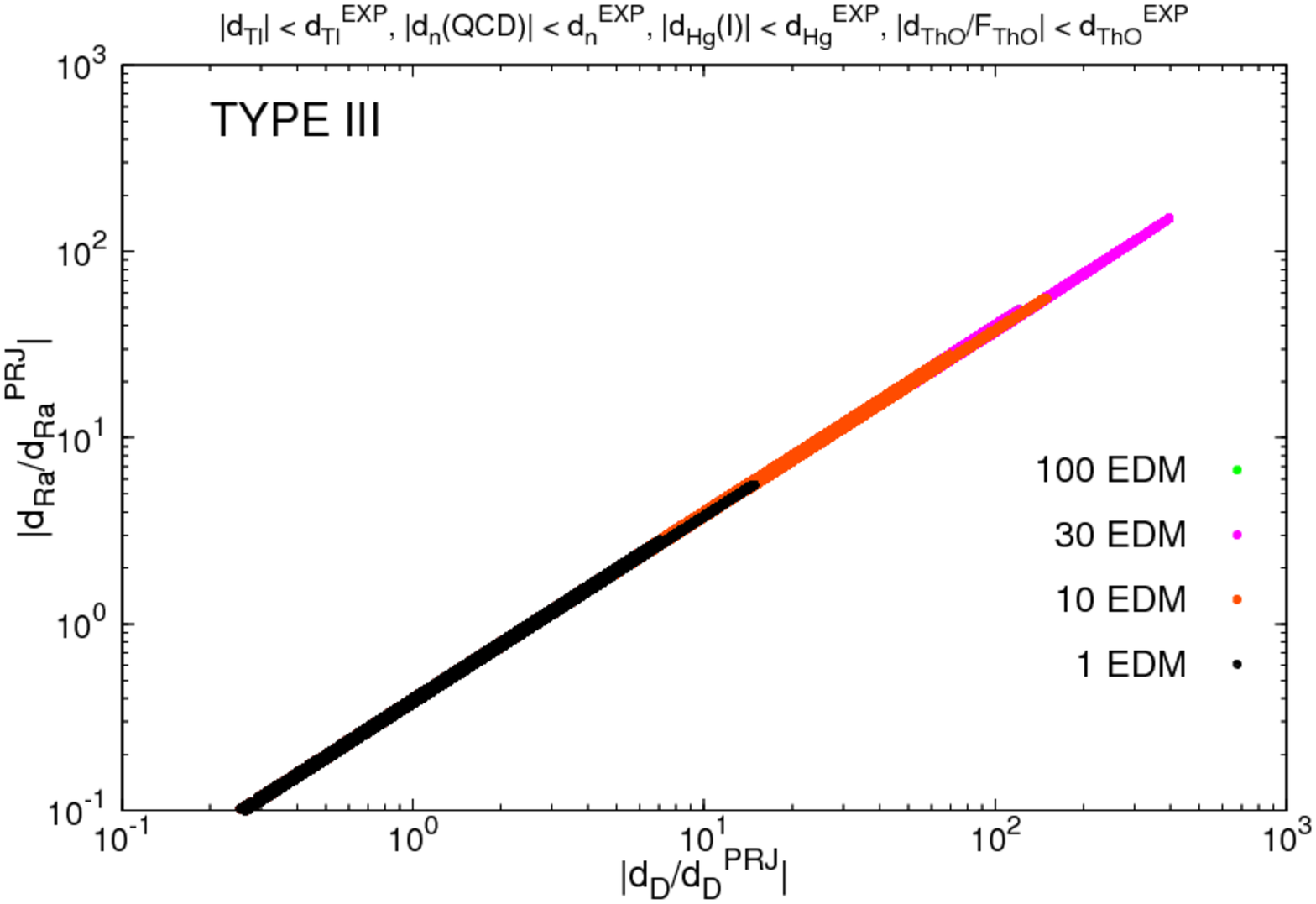}
\includegraphics[angle=-90,width=7cm]{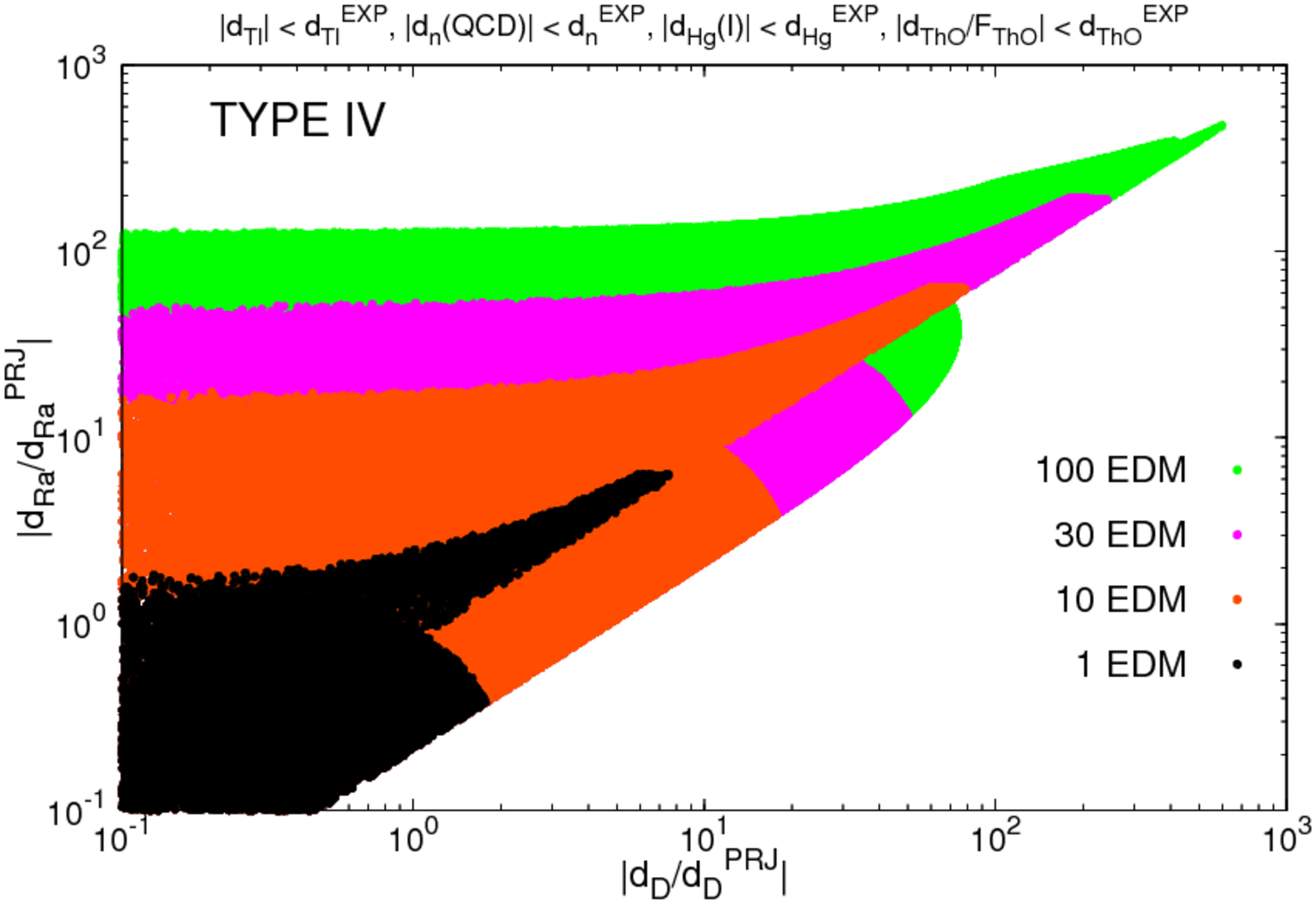}
\vspace{0.5cm}
\caption{\it The correlation between 
$|d_{\rm D}/d_{\rm D}^{\rm PRJ}|$ and
$|d_{\rm Ra}/d_{\rm Ra}^{\rm PRJ}|$
in the $95\,\%$ CL regions satisfying
the Thallium, thorium-monoxide, neutron, and Mercury EDM constraints 
simultaneously
taking the relaxation factor $r=1$ (black),
$r=10$ (orange), $30$ (pink), and $100$ (green).
}
\label{fig:ddra}
\end{figure}
Finally, in Fig.~\ref{fig:ddra}, we show the correlation between 
$|d_{\rm D}/d_{\rm D}^{\rm PRJ}|$ and
$|d_{\rm Ra}/d_{\rm Ra}^{\rm PRJ}|$ in the colored regions of
Fig.~\ref{fig:hfit_95} with  $r=1$ (black),
$r=10$ (orange), $30$ (pink), and $100$ (green).
Note that the projected sensitivities for the deuteron EDM $d_D^{\rm PRJ}$ and
the Radium EDM $d^{\rm PRJ}_{Ra}$ can be found right after Eq.~(\ref{eq:exp}).
The strong correlations seen in Types I and III can be understood by observing 
that the dominant contributions to $d_{\rm D}$ and $d_{\rm Ra}$ coming
from $d^C_{u,d}$ and $d^G$ are all proportional to the product $C_u^S\times C_u^P$
with no dependence on $C_v$, see Figs.~\ref{fig:dcu}, \ref{fig:dcd}, and
\ref{fig:dgw}.
The ratios $|d_{\rm D}/d_{\rm D}^{\rm PRJ}|$ and $|d_{\rm Ra}/d_{\rm Ra}^{\rm
PRJ}|$ lying in the ranges from about $10$ and $100$ 
require the degree of cancellation of $90\,\%$ (orange regions).
Even in the black regions ($r=1$) without any additional contributions beyond
those from the $125.5$ GeV Higgs, we find that the deuteron EDM can be 
$5$ (I), $10$ (II), $15$ (III),
and $8$ (IV) times as large as the projected experimental sensitivity.
While those for the Radium EDM are
$2$ (I), $7$ (II), $6$ (III), and $7$ (IV)
times as large as the experimental sensitivity.
It means that  
the deuteron and Radium EDMs can be easily above the projected
sensitivities offered by the new experiments
even when the combined EDM constraints are the most stringent
without assuming any
additional contributions beyond those from the 125.5
GeV Higgs boson.

\section{Conclusions}

In this work, we have updated the Higgcision constraints on the Higgs
boson couplings to SM gauge bosons and fermions, and confronted
the allowed parameter space in $C_u^S$, $C_u^P$, and $C_v$ against
various EDM constraints from the non-observation  of  the 
Thallium ($^{205}{\rm Tl}$), thorium-monoxide (ThO),
neutron, and Mercury ($^{199}{\rm Hg}$) EDMs,
in the framework of 2HDMs.
Although the Higgs boson data still allow sizable $C_u^P$, the 
combined EDM constraints restrict $|C_u^P|$ to a very small value
of $\sim 10^{-2}$. 

We have only considered the contributions from the 125.5 GeV Higgs boson
via the Higgs-mediated diagrams in this work. 
There could potentially be contributions
from other particles of any new physics models, e.g., the heavier Higgs bosons
of multi-Higgs models, supersymmetric particles, or any other exotic 
particles that carry CP-violating couplings. 
These contributions and the contributions
from the 125.5 GeV Higgs boson could cancel each other in a 
delicate way.  If we allow 1\% fine tuning, the constraints on the
pseudoscalar coupling $C_u^P$ are relaxed and $|C_u^P|$ as large as $0.5$ 
can be allowed.

\bigskip

In the following we offer a few more comments before we close.

\begin{enumerate}
\item The observable EDMs involve the electron EDM $d^E_e$,
(C)EDMs of the up and down quarks $d^{E,C}_{u,d}$, and
the coefficient of the Weinberg operator $d^{\,G}$.
Only $d^{\,G}$ 
is independent of the Higgs couplings to the first-generation fermions.

\item The observed $125.5$ Higgs boson, which is denoted as $H$ in this work,
gives definite predictions
for $d^E_e$ and $d^{E,C}_{u,d}$
through the two-loop Barr--Zee diagrams. 

\item
For $d^E_e$, we consider both the Barr--Zee diagrams
mediated by the $\gamma$-$\gamma$-$H$ couplings and
by the $\gamma$-$H$-$Z$ couplings with 
the constituent contributions from top, bottom, 
tau, and W-boson loops. We note the $\gamma$-$\gamma$-$H$ 
Barr--Zee diagrams are dominant.
We further observe that
the contributions from top and W-boson loops are dominant and 
a cancellation occurs between them
around $C_v=1$ in Types II and III. Note 
the current Higgs data prefer the region around $C_v=1$. 

\item For $d^E_{u,d}$, the contribution from the $\gamma$-$\gamma$-$H$
and $\gamma$-$H$-$Z$ Barr--Zee diagrams are comparable. In $d^E_d$,
a cancellation occurs between them
around $C_v=1$ in Types II and IV.

\item The Barr--Zee contributions to $d^C_{u,d}$
are dominated by the top loops which are
independent of the 2HDM types except for $d^C_d$ in Types II and IV.

\item The dominant contributions to 
$d^{\,G}$ from top loops are independent of the 2HDM types.

\item The Thallium and ThO EDMs are dominated by $d^E_e$,
the neutron EDM by $d^C_{u,d}$ and $d^G$, and the
Mercury EDM by $d^C_{u,d}$ through the Schiff moment. 
We observe a cancellation occurs between the contributions from $d^C_{u,d}$ and
$d^G$ to the neutron EDM
around $C_v=1$ in Types II and IV.

\item The ThO (neutron) EDM constraint is relatively weaker
in Types II and III (Types II and IV), while
the Mercury EDM constraint
is almost equally stringent in all four types.

\item We find that
the deuteron and Radium EDMs can be $\sim 10$ times as large as the projected
experimental sensitivities
even when $|C_u^P|$ is restricted to be smaller than about $10^{-2}$
by the combined EDM constraints.

\end{enumerate}

{\bf Note Added}:
After the completion of this work, we received a paper~\cite{4257},
which addresses the LHC Higgs and EDM constraints in Types I and II 2HDMs.

\section*{Acknowledgment}  
We thank Wouter Dekens and Jordy de Vries for helpful discussions and
valuable comments.
This work was supported the National Science
Council of Taiwan under Grants No. 102-2112-M-007-015-MY3.
J.S.L. was supported by 
the National Research Foundation of Korea (NRF) grant 
(No. 2013R1A2A2A01015406).
This study was also
financially supported by Chonnam National University, 2012.


\end{document}